\documentclass[12pt]{iopart}
\usepackage{graphicx}
\usepackage{deluxetable}

\begin{document}

\title[]{Toward the general RGB slope-metallicity-age calibration: I Metallicities, Ages and Kinematics for Eight LMC Clusters \footnote{Based on observations collected with the FORS2-VLT of the ESO  within observing program 078.D-0428(B).}}

\author{Saurabh Sharma$^1$, J. Borissova$^1$, R. Kurtev$^1$, V. D. Ivanov$^2$ and  D. Geisler$^3$ }

\address{$^1$ Departamento de F\'isica y Astronom\'ia, Facultad de Ciencias, Universidad de Valpara\'{\i}so,
       Ave. Gran Breta\~na 1111, Playa Ancha, Casilla 53,
             Valpara\'iso, Chile
\ead{saurabh@dfa.uv.cl;jura.borissova@uv.cl; radostin.kurtev@uv.cl} }

\address{$^2$ European Southern Observatory, Ave.\ Alonso de Cordoba 3107,
        Casilla 19, Santiago 19001, Chile \ead{vivanov@eso.org} }

\address{$^3$ Departamento de Astronom\'{\i}a,  
Universidad de Concepci\'{o}n, Casilla 160-C, Concepci\'{o}n, Chile \ead{dgeisler@astro-udec.cl} }

\begin{abstract}
In this paper, we discuss the properties of CMDs, age, metallicity and 
radial velocities of eight massive LMC  clusters 
using data taken from FORS2 multiobject spectrograph at the 8.2-meter VLT/UT1.
The strong  near-infrared Ca II triplet (CaT) lines of RGB stars 
obtained from the high S/N spectra  are used to 
determine the metallicity and radial velocity of cluster members.
We report for the first time spectroscopically determined metallicity values 
for four clusters based on the mean [Fe/H] value of $\sim$10 cluster members each.
We found two concentrations in the distribution of ages of the target clusters. Six have ages between 
0.8-2.2 Gyr and the other two, NGC 1754 and NGC 1786, are very old.
The metallicity of the six intermediate age clusters, with   a 
mean age of 1.5 Gyr, is $-0.49$ with a scatter of only
0.04. This tight distribution suggests that a close encounter between the LMC and SMC may have 
caused not only the restart of cluster formation in the LMC but the generation of the central bar.
The metallicity for the two old clusters is similar to that of the other old, metal-poor  LMC clusters.
We find that the LMC cluster system exhibits disk-like rotation with no clusters appearing to have halo kinematics
and there is no evidence of a metallicity gradient in the LMC, in contrast with the stellar
population of the MW and M33, where  the metallicity decreases as galactocentric distance increases.
The LMC's stellar bar may be the factor responsible for the dilution of any kind of gradient in the LMC.
\end{abstract}


\section{Introduction}

The  Large Magellanic cloud (LMC) is an important target in the effort to understand the stellar populations of galaxies, due to its close
proximity, favorable viewing angle, ongoing star-formation activity (Harris \& Zaritsky, 2009) and relatively low metallicity.
The LMC cluster system is well known to show a puzzling age distribution, with a handful of old ($\sim$13 Gyr),
metal-poor globular clusters and a number of intermediate-age (1-3 Gyr), relatively metal-rich populous clusters, with only a single cluster  `ESO 121-SC03' in between (Mackey et al. 2006 and Mateo et al. 1986).
The chemical enrichment of the LMC has been studied in some detail recently by
Harris \& Zaritsky (2009) and Carrera et al. (2008). 
The early history of the LMC was characterized by rapid enrichment, 
reaching [Fe/H]=$-1.2$ by about 12 Gyr ago  (Harris \& Zaritsky, 2009). The metallicity then increased only 
slightly throughout the long quiescent epoch. When cluster formation resumed some 
3 Gyr ago, the chemical abundances began a steady ascent to the present-day value of 
 [Fe/H]=$-0.2$ - $-0.3$ (Carrera et al. 2008; Hill et al. 1995; and  Rolleston, Trundle \& Dufton 2002). 

These features provide clues to the history of star formation in the LMC.
There was an initial epoch of star formation during which a significant portion of the 
LMC’s stellar mass was formed. This era was followed by a long quiescent epoch, during which cluster
formation was suppressed throughout the LMC. Then, about 3 Gyr ago, cluster formation resumed throughout the
LMC, and has been ongoing since. 
The global resumption of star formation is strongly suggestive of a dramatic
change at this point in the LMC's past, such as the merger of a gas-rich dwarf galaxy or a particularly dramatic tidal
encounter, probably with the Small Magellanic cloud (SMC) given the similar resumption of star formation seen in 
that galaxy’s star formation history (SFH) (Harris \& Zaritsky, 2009).
Bekki et al. (2004) suggest that the recent burst of cluster formation is linked
to the first very close encounter between the Clouds about 4 Gyr
ago, which would have induced dramatic gas cloud collisions,
allowing the LMC to begin a new epoch of cluster and star formation; 
strong tidal interactions between the Clouds have likely
sustained the enhanced cluster formation. Bekki et al. (2004)
also find that the close encounter between the LMC and SMC would have
been sufficient to cause the formation of the LMC Bar around
the time of the new epoch of cluster formation, giving rise to the
similar SFHs seen in the cluster system and the bar (Cole et al. 2005, hereafter C05).
In addition to enhancing star formation, tidal forces can result in the in-fall or
outflow of material, thereby affecting the chemical evolution history (CEH) of the LMC and, at
the same time, leaving behind a signature of the interaction. Thus,
accurate knowledge of the ages and metallicities of LMC clusters
is necessary to fully understand the formation and dynamical history of this galaxy.
Here it is worthwhile to mention that recently Kallivayalil  et al. (2006, 2009), using HST data for fields in
the Magellanic Clouds (MCs) centered on background quasars, determined the
proper motions of the LMC and SMC. They argued that the 
MCs may not be bound to each other or the Milky Way (MW) and may in fact be passing the MW for the first time,
although, to further verify and refine these results, more precise data is needed.

Red giant branch (RGB) stars are among the brightest red stars in stellar systems
that are older than $\sim$ a Gyr. They appear in almost all galaxies. Therefore, red giants are an excellent tool
for probing the parameters of old populations and the history of star formation in almost any galaxy. 
Globular clusters, with their single age and metallicity, are the ideal sites for calibrating the RGB parameters. Since
Da Costa \& Armandroff (1990) explored the possibility to use the color of the RGB as a
metallicity indicator, many signiﬁcant advances have been made, both in the improvement of the astronomical
instrumentation and the development of the corresponding theory.
Kuchinski et al. (1995) and Minniti et al. (1995) demonstrated both empirically
and theoretically that the slope of the RGB in a $K$ vs. $(J-K)$ color-magnitude diagram (CMD) is sensitive to the
metallicity of the population. They investigated this correlation for metal-rich globular clusters and derived a
linear relation between [Fe/H] and RGB slope. Ivanov et al. (2000) extended this relation to
metallicities as low as [Fe/H] = $-2.1$. Ferraro et al. (2000) derived a similar relation. Ivanov \& 
Borissova (2002) tabulated the relation in [$M_K$, $(J- K_S)$, [Fe/H]] space based on a large sample
of 2MASS photometry of MW globular clusters.
Tiede et al. (1997), using 4 Galactic open clusters, derived the slope of the RGB - metallicity
relation for a younger population and Kyeong \& Byun (2001) re-derived the calibration adding
10 new Galactic open clusters. And finally, Tiede et al. (2002) found two very clear linear
trends, one for Galactic open clusters spanning $-0.6<$ [Fe/H] $< +0.1$ and another for Galactic globular clusters
spanning $-2.2 <$ [Fe/H] $< -0.4$.

To derive the general relation between
slope of the RGB, [Fe/H] and age, we have to extend the existing calibrations to the clusters in
external galaxies, because there are no Galactic counterparts of the young metal-poor MC clusters.
Accordingly, we observed more than a dozen clusters in the LMC and SMC in December, 2004 with the WFIRC
attached to the 2.5-m du Pont telescope in the Las Campanas observatory. 
Thus, we have very good-quality CMDs, which
allow us to determine the slope of the RGB of the clusters. The ages of the clusters are determined from good-
quality HST CMDs with well-defined main-sequence turn-off points. 
The metallicities of most clusters in our RGB slope
sample are based on the latest spectroscopic measurements available
in the literature (G06). There are eight clusters given in Table \ref{log}, however, with
no modern metallicity determination. 
There is an empirically developed, simply calibrated method available which can make 
an efficient estimate of metallicity ([Fe/H]) for individual RGB 
stars using the strength of the Ca II triplet (CaT) lines at 8498, 8542, and 8662 \AA
(Battaglia et al. 2008; Carrera et al. 2007;  Cole et al. 2004, hereafter C04; C05;
Grocholski 2006, hereafter G06; Parisi 2009, hereafter P09; Warren \& Cole 2009 etc.).
A powerful application of this method to the age-metallicity relation and abundance gradient of the star
clusters of the LMC and SMC were already demonstrated by G06 and P09, following up on the pioneering 
work by Olszewski et al. (1991, hereafter O91). Therefore,
to derive the metallicity, we obtained medium-resolution spectra at the near-infrared CaT
using the FORS2 multiobject spectrograph at the 8.2-meter VLT/UT1 with high signal-to-noise ($20 \leq S/N \leq 100$) ratio.
The benefit of using CaT is that it is very
efficient; not only are RGB stars near their brightest in the infrared, but the multiplexing
capabilities of many moderate-resolution spectrograph's allow the observation of dozens of
stars simultaneously, greatly increasing the probability of identifying cluster members (P09). 
In addition, the metallicity calibration is very straightforward and well
calibrated and independent of luminosity, distance, reddening and age.
Therefore, using FORS2 on the VLT, we were able to identify 77 member stars
in eight massive LMC clusters and determine accurate mean cluster velocities 
and metallicities.

In this paper, we discuss the metallicity distribution, their relationship and the kinematics
of LMC clusters, extending the work done by G06. We will further discuss the relation between
RGB slope, metallicity and age in an upcoming paper (Kurtev et al. 2010).
In \S2, \S3, \S4, and \S5, we discuss the observations, techniques required to determine 
radial velocity \& metallicity and the cluster membership criteria.
In \S6 \& \S7, we present   the basic cluster
parameters based on our photometry and spectroscopy and
analyze the results, followed by conclusions in \S8.

\section{Observations and data reduction}

\subsection{Photometry}

Short exposure images of eight LMC clusters were observed
with FORS2, VLT, through the filters $V$ and $R$ under excellent atmospheric conditions (seeing $<$ 0.7 arcsec) 
on 5 October 2006 and 17 November 2006. FORS2 was used in normal
resolution with the pixel size of 0.25$''$/pixel and field of view of $6.8^\prime\times6.8^\prime$,
which allowed us to sample, in the outermost field regions, a good portion of
the background LMC field.  The coordinates of the clusters as well as the total exposure times are listed in Table \ref{log}.

De-biasing and flat-fielding were performed with standard IRAF
packages, using the calibration frames (bias and sky flat fields)
obtained during the same nights as the target observations.
Photometric reduction was carried out with the
DAOPHOT~II/ALLFRAME package (Stetson 1987, 1994). Stellar PSF models
for each frame were obtained by means of a large set of uncrowded and
unsaturated stars. The mean photometric errors are 0.003-0.009 mag for 
($V, R$) $<$ 20 mag and between 0.01 mag and 0.03 mag for the fainter magnitudes. 
The nights were photometric and the calibration to the Johnson
standard photometric system was performed by comparison of the stars in the 
Landolt (1992) field PG2213-006. The maximum zero-point errors are calculated as $\pm$ 0.01-0.02 magnitudes.

\subsection{Spectroscopy}

The spectroscopic observations were carried out with FORS2
in visitor mode at the Antu (VLT-UT1) 8.2 m telescope at ESO’s
Paranal Observatory during the night of 16-17 November 2006. 
We used the FORS2 spectrograph in mask exchange unit (MXU) mode,
with the 1028z+29 grism and OG590+32 order blocking filter.
The MXU slit mask configuration allows the placement of more
slits (in this case 37) on the sky than the 19 movable slits provided in Multi Object 
Spectrograph mode.  Targets were
selected so as to maximize the number of likely cluster members
observed; typically 20 stars inside our estimated cluster radius (see \S5)
were observed, with an additional 15 stars outside of this radius
that appeared to be LMC field red giants based on our infrared photometry with 
the du Pont telescope.  

We used slits that were 1$^{\prime\prime}$ wide and 6.25$^{\prime\prime}$ long. 
In all fields, we obtained 3 to 4 separate 600s exposures, taken under seeing less than $1^{\prime\prime}$. They were average combined to form the final spectra.
Pixels were binned $2\times2$, yielding a plate scale of 0.25$^{\prime\prime}$ pixel$^{-1}$. 
We used the upper (master CCD) of FORS2 for observations of the target clusters  
which has a readout noise of 2.9 $e^-$ and inverse gain of 0.7 $e^-$ ADU$^{-1}$.
The spectra have a dispersion of $\sim 0.85$ \AA/pixel
(which corresponds to a resolution of 2-3 \AA) with a characteristic rms scatter of ∼ 0.06 \AA,
and cover a range of $\sim 1600$ \AA ~in the region of the CaT (8498, 8542, and 8662 \AA). 
For the final spectra, S/N is typically 40-60 pixel$^{-1}$ with some stars as high as $~\sim 100$ pixel$^{-1}$
and, in some few cases, as low as  $~\sim 20$ pixel$^{-1}$. For the present study we use spectra having S/N $\geq 35$ pixel$^{-1}$.
Calibration exposures, bias frames and flat-fields were also taken as a part of the standard calibration plan.

\begin{figure*}
\centering
{\includegraphics[height=7cm,width=7cm,angle=0]{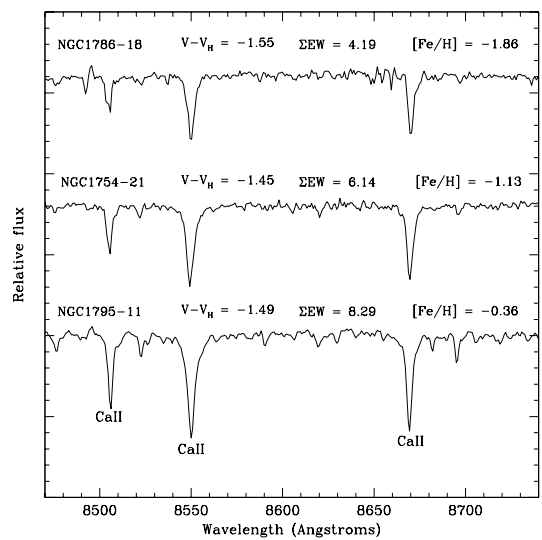}}
\caption{Sample spectra from RGB stars in our target clusters covering a
range in metallicities. The three Ca II lines are
marked for reference; the change in CaT line strength with [Fe/H] is readily
visible. Calculated summed equivalent widths and metallicities for each star are given.}
\label{spec}
\end{figure*}

We used the IRAF task ccdproc to fit and subtract the over-scan region, trim the images, 
and flat-field each image. Flats were corrected for the lamp shape by the task response. 
We used the task apall to
define the sky background and extract the stellar spectra onto one dimension.
The sky level was defined by performing a linear fit across the dispersion direction 
to sky windows on each side of the star. When the star fell near the top or bottom of 
the slitlet; the sky regions were chosen interactively.
The tasks identify and dispcor were used to calculate and apply the dispersion solution for
each spectrum. Finally, the spectra were continuum normalized by fitting a polynomial to
the stellar continuum. Examples of the final spectra in the CaT region can be seen in
Figure \ref{spec}. Clearly, the three Ca II lines dominate the red giant’s spectra in this region. Also
note the stronger Ca lines in the bottom spectrum, which, given the very similar $T_{eff}$ and
log $g$ values (as indicated by their virtually identical $V-V_{HB}$ values, i.e.
the mag difference between the star and the horizontal branch), graphically illustrates
the higher metallicity of the bottom spectrum and the power of the CaT technique.

\section {Radial Velocity }

Radial velocity measurements for our target stars has been done
to help us in defining the cluster membership as well as to study the kinematics of the LMC.
Since a cluster’s velocity dispersion is expected to be relatively small 
compared to the surrounding field and its mean velocity quite possibly distinct from the
field, radial velocities are an excellent tool for determining cluster membership. 

Radial velocities for all the target stars were determined through
cross-correlation with eight template spectra from NGC 1851 (Carerra et al. 2007) 
using the IRAF task fxcor (Tonry \& Davis 1979). These template stars were observed as a part of the 
CaT calibration program; thus, this sample offers a good
match to the spectral types of our target stars. In addition, these
observations were made with a telescope and instrument setup
that is almost identical to those used by us. 
In addition to calculating relative radial velocities, fxcor
uses information about the observatory location, date and the
time of the observations (once the ESO header has been appropriately reformatted) 
to correct the derived velocities to the heliocentric reference frame. 
For a star’s final heliocentric radial velocity, we adopt the mean  of all
cross-correlation results. We find good agreement among the template-derived
velocities, with a typical standard deviation of 8 km s$^{-1}$.
Having used the same set up as G06, we follow the same approach to
calculate the error due to imprecise alignment of the slit center and the stellar centroid.
G06 estimate this latter error as $\pm4.2$ km $s^{-1}$. This, together
with the the standard deviation of 8 km s$^{-1}$
gives a typical error of 9  km s$^{-1}$ in measuring the radial velocity in the
present study.

\section {Equivalent widths and Abundances}

The abundances of the clusters in the present sample have been derived from the equivalent 
width (EW) of strong CaT lines in individual RGB stars based on the empirically developed
method which has been explored already in detail previously (see for eg. C04, C05, G06, P06).
The CaT lines can be contaminated by weak neutral metal lines, and the continuum
may be affected by weak molecular bands. Both effects make it impossible 
to measure the `true' EWs of the CaT lines in isolation. 
Armandroff \& Zinn (1988) defined two pseudocontinum windows in line free regions on either side of each feature. 
For each of the Ca II lines, the line connecting the average number of the counts in each of these 
two windows was defined to be the pseudocontinum. The pseudo-equivalent width was then computed by 
taking the integral over the feature boundaries of the difference between the pseudocontinum and the spectrum.
Further, C04 mentioned that the spectra need not be flux-calibrated because the continuum 
slopes of red giants in this wavelength range are generally close
to flat, so a simple normalization by an arbitrary polynomial suffices 
for measurement of the ‘pseudo-equivalent width’. 
In the present study, we calculate  the EW by direct pixel-to-pixel integration 
using the interactive splot program of IRAF.
The EWs are measured by visually marking two continuum points at the line band pass given by Armandroff \& Zinn (1988).
The continuum level,  which is a linear function between these two marked points, is effectively fixed at 1 by our normalization procedure.
The integral is taken between these two points.  
Since our spectra are of high enough quality that all three CaT lines are well measured,
we define our metallicity index $\Sigma W$, as the sum of the  EWs of three CaT lines (see for eg.  C04, G06, and P09).

Although theoretical and empirical studies have shown that effective temperature,
surface gravity, and [Fe/H] all play a role in CaT line strengths (e.g., Jorgensen et al.
1992; Cenarro et al. 2002), Armandroff \& Da Costa (1991) showed that there is a linear
relationship between a star’s absolute magnitude and $\Sigma W$ for red giants of a given
metallicity. Following previous authors using the same CaT technique (G06 \& P09), we define a reduced equivalent width, 
W$^\prime$, to remove the effects of surface gravity and temperature on $\Sigma W$ via its luminosity dependence:

                                  $W^\prime   = \Sigma W + \beta(V - V_{HB}),$   

The introduction of the difference between the visual magnitude of the star ($V$) and
the cluster’s horizontal branch/red clump $(V_{HB} )$ in calculating W$^\prime$, also removes any dependence on cluster
distance and interstellar reddening. The value of $\beta$ has been investigated in detail by previous
authors (C04, G06, P09) and we have adopted the value obtained by C04: $\beta =  0.73 \pm 0.04$. We chose
C04’s value of $\beta$ because they used both globular and open clusters, covering a broad range of
ages, to derive it and their instrumental setup is very
similar to ours. We note that, as discussed in detail in G06, the use of C04’s value for $\beta$,
combined with defining the brightness of our target RGB stars relative to the HB of their
parent cluster, incorporate any age effects into the CaT calibration, thus it is not necessary
to make any corrections for the ages of our clusters. 

As shown by Rutledge et al. (1997),
there is a linear relationship between a cluster’s reduced EW and its metallicity on the
Carretta \& Gratton (1997) abundance scale for globular clusters of the MW. C04
extended this calibration to a wider range of ages ($2.5$ Gyr $\leq$ age $\leq$ 13 Gyr) and metallicities
$(-2.0 \leq$ [Fe/H] $\leq -0.2)$ by combining the metallicity scales of Carretta \& Gratton (1997)
and Friel et al. (2002) for globular and open clusters, respectively. Although many of our
clusters are younger than 2.5 Gyr, we adopted the C04 relationship,

                       [Fe/H] = $(-2.966 \pm 0.032) + (0.362 \pm 0.014)W^\prime$,

to derive the metallicities of our entire cluster sample. Carrera et al. (2007) investigated
the behavior of CaT in the age range of 0.25 Gyr $\leq$ age $\leq$ 13 Gyr and the metallicity range
$-2.2 \leq$ [Fe/H] $\leq +0.47$. In spite of the extended age range of their calibration, we decided
to use C04’s calibration for three reasons. First, the calibration of Carrera et al. (2007)
uses the absolute magnitude of a target star (rather than the brightness relative to the HB,
as we have adopted), thus requiring accurate photometric calibration and distance to the
star. Second, using $(V - V_{HB} )$ instead of absolute magnitude removes the
need to make any assumptions about the foreground reddening toward a cluster. Finally,
Carrera et al. (2007) performed estimates of the differences in W$^\prime$ as a function of age using
the Jorgensen et al. (1992) models and BaSTI stellar evolution models (Pietrinferni et al.
2004). From their data, they confirmed that the influence of age is small (for clusters having age
10.5 and 0.6 Gyr, the difference in metallicity is $\sim 0.25-0.15$ dex), even for ages $<$ 1 Gyr. 
Based on these results, we are confident that the C04 calibration can be accurately
extended to younger ages (see also P09). 

     In addition, Battaglia et al. (2007) have shown that [Fe/H] values derived from CaT
agree with those from high resolution spectroscopy and detailed model atmosphere analysis
to within 0.1 - 0.2 dex over the range $-2.5 \leq$ [Fe/H] $\leq -0.5$, based on FLAMES data for a
large sample of stars in the Sculptor and Fornax dwarf spheroidal galaxies. It is well known
that such stars have distinct [Ca/Fe] ratios from those in the Galactic globular cluster
calibrators (e.g., Shetrone et al. 2001; Geisler et al. 2007) and also possess a larger range
of stellar ages, extending to much smaller values. Nevertheless, CaT yields metallicities
that are very close to real Fe abundances, reinforcing our confidence in the technique. 

With the assumption that each star in a
given cluster has the same metallicity and thus the same $W^\prime$ value,
the error on $W^\prime$, $\sigma_W{^\prime}$, depends on the standard deviation of the
individual star's $W^\prime$ values about the mean cluster value as well as
the number of stars, N, used in each cluster (see also Warren \& Cole 2009) i.e.
${\sigma_W{^\prime} }= {\sigma \over \sqrt{ (N)} }$, this error is further propagated 
to define errors in calculating the [Fe/H]. Our
estimate of the systematic error per star ranges from 0.08 to 0.14 dex, with a mean of 0.11 dex.

\section {Cluster Membership}

\begin{figure*}
\centering
\hbox{
\includegraphics[height=6cm,width=6cm,angle=-90]{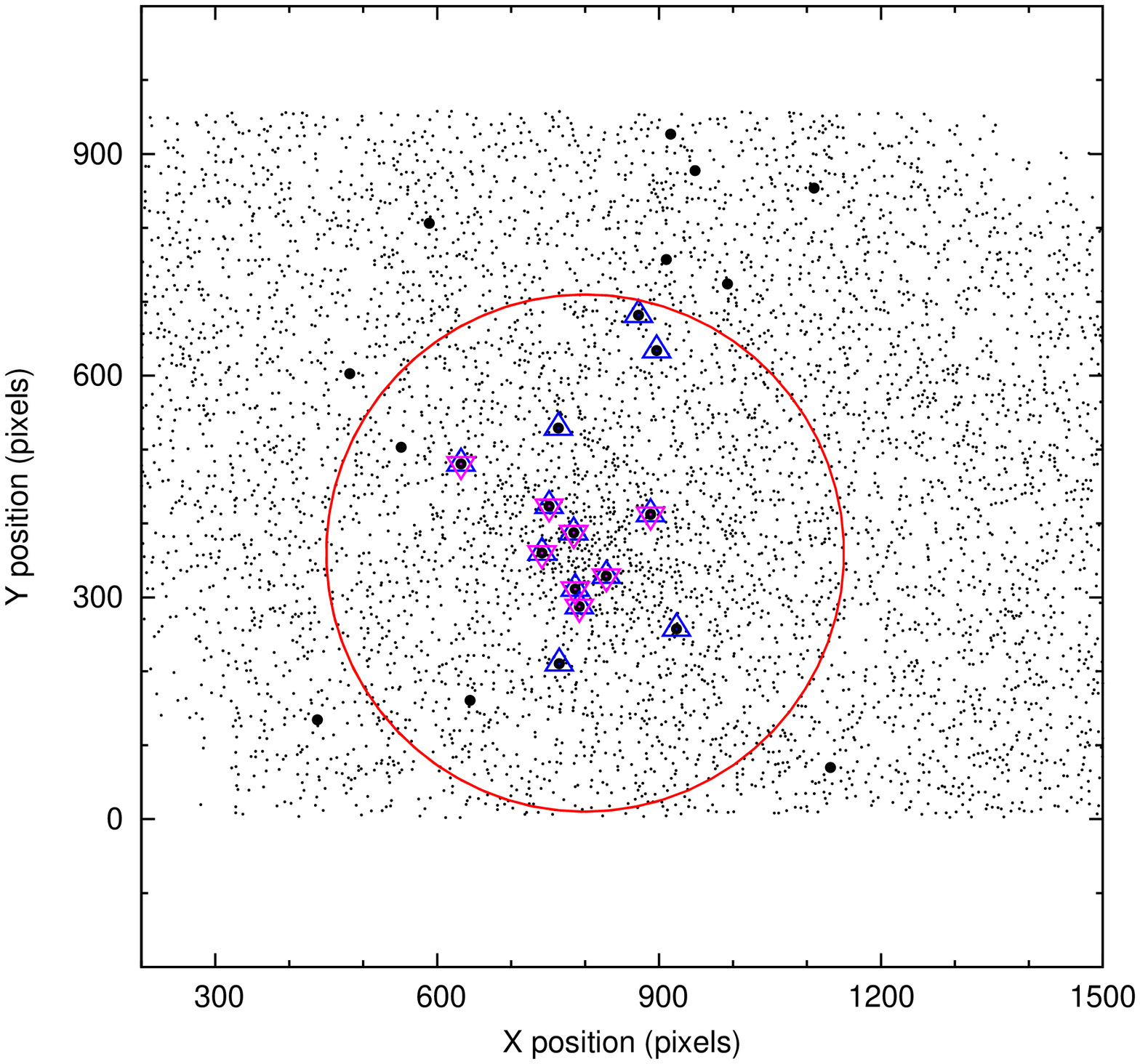}
\includegraphics[height=8cm,width=6cm,angle=-90]{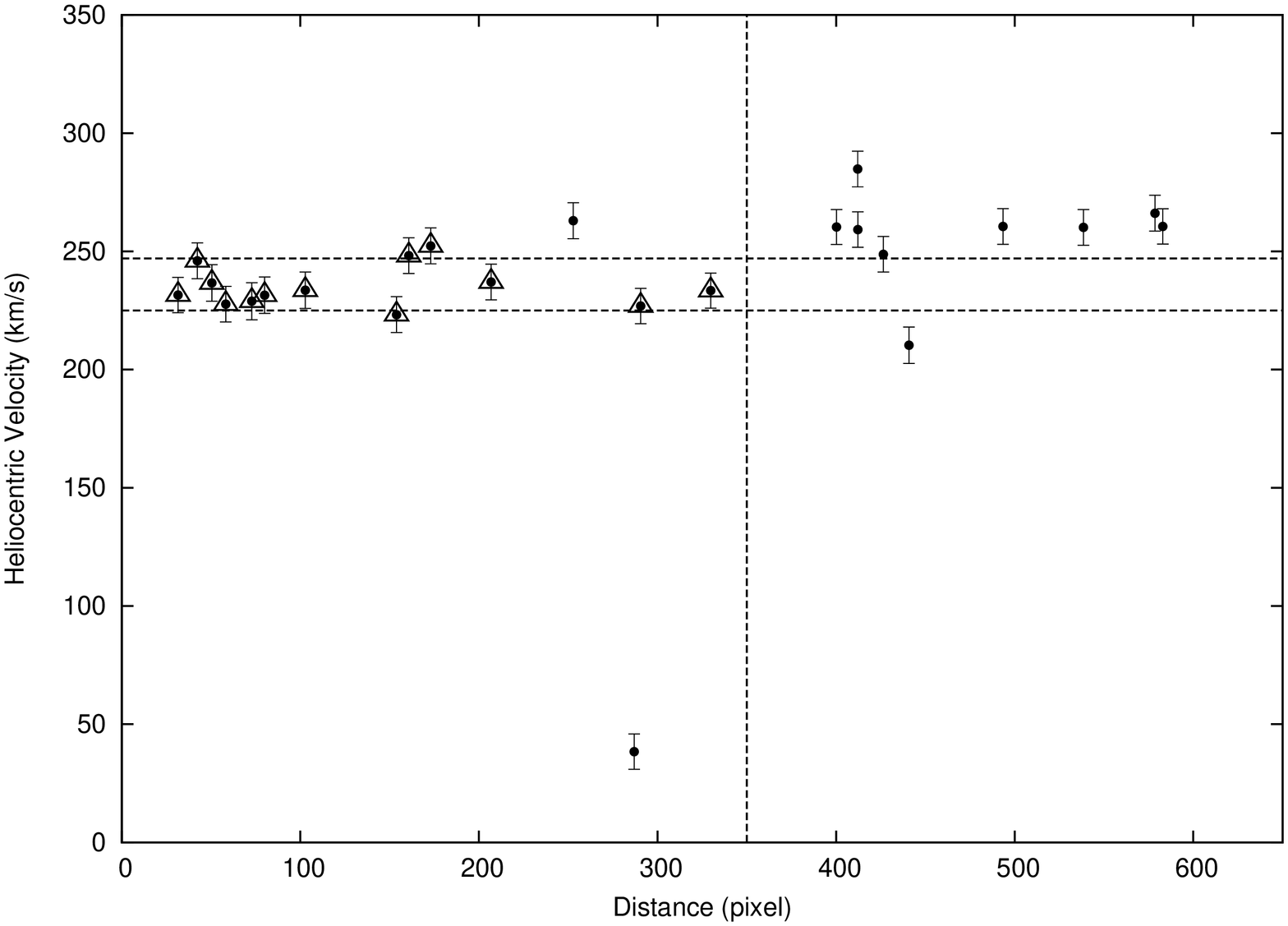}
}
\hbox{
\includegraphics[height=6cm,width=6cm,angle=-90]{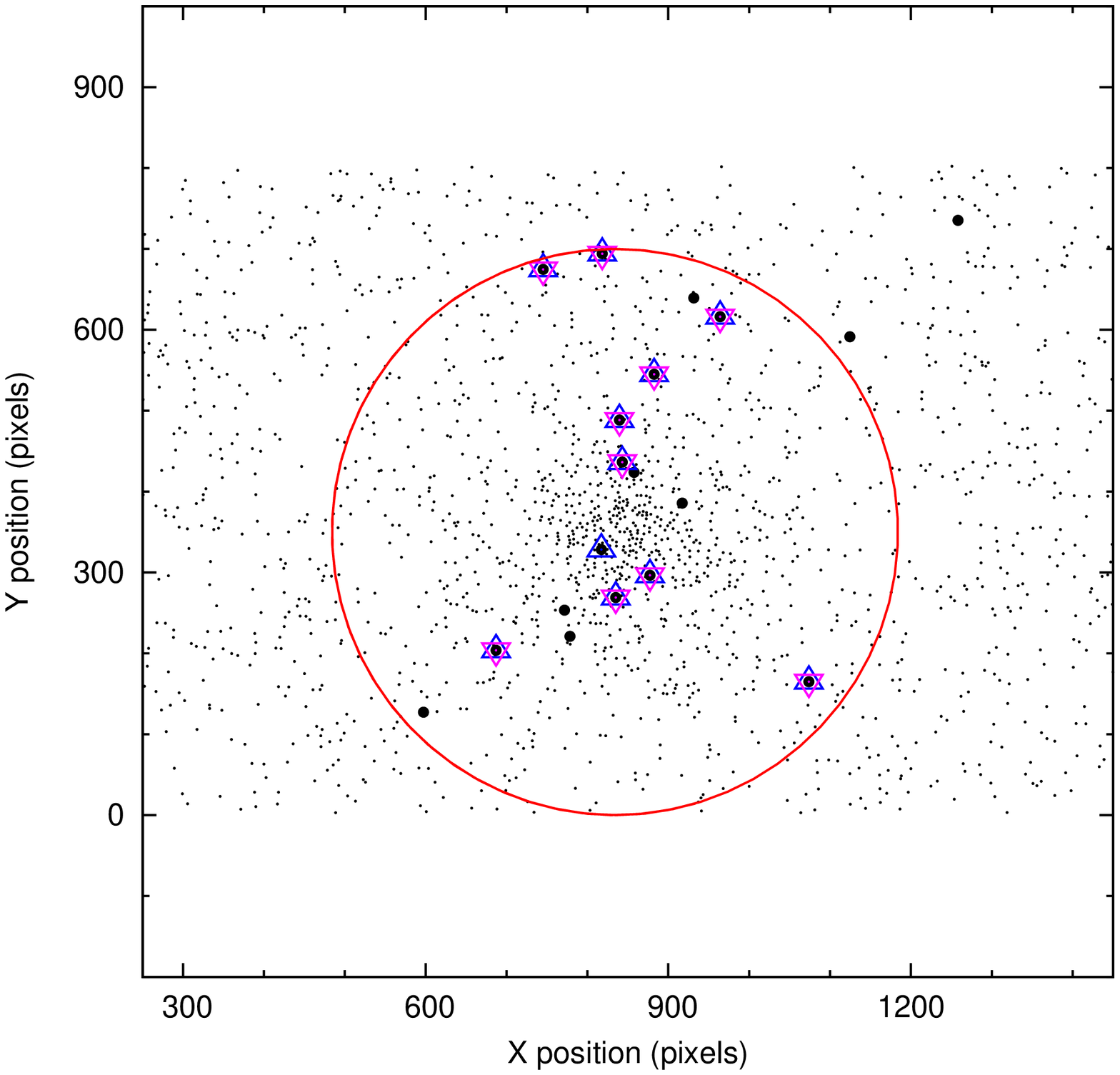}
\includegraphics[height=8cm,width=6cm,angle=-90]{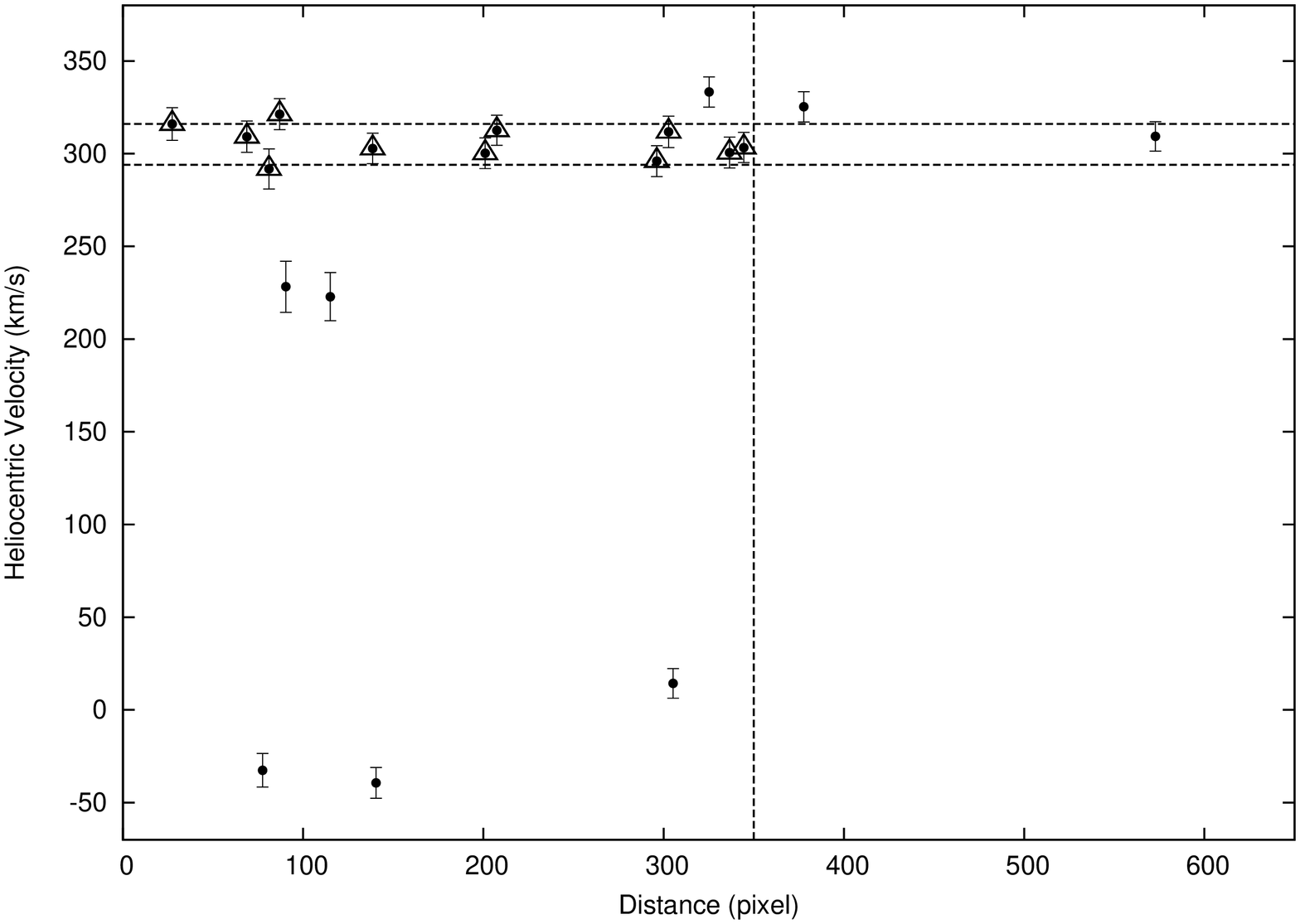}
}
\caption{
({\it left  panels}) The (x, y) positions for the stars in cluster NGC 1754 ({\it upper}) and NGC 1900 ({\it lower}).
The adopted cluster radius is marked by the large open circle,
and stars outside this radius are considered non-members. 
The blue color triangles are the stars which are selected from the radial velocity
 criterion                      
and then the final cluster members (purple inverse triangle) from these blue stars have been selected on the basis of metallicities
(see text for details).
({\it right panels}) Radial velocities for our spectroscopic targets as a function of distance 
from the  NGC 1754 ({\it upper}) and NGC 1900 ({\it lower}) cluster centers. 
The horizontal lines represent our velocity cut and have a width of $\pm11$ km $s^{-1}$. 
The cluster radius is shown by the vertical line. The error bars represent the
random error in determining the radial velocity for each star.
}
\label{map}
\end{figure*}

\begin{figure*}
\centering
\hbox{
\includegraphics[height=8cm,width=6cm,angle=-90]{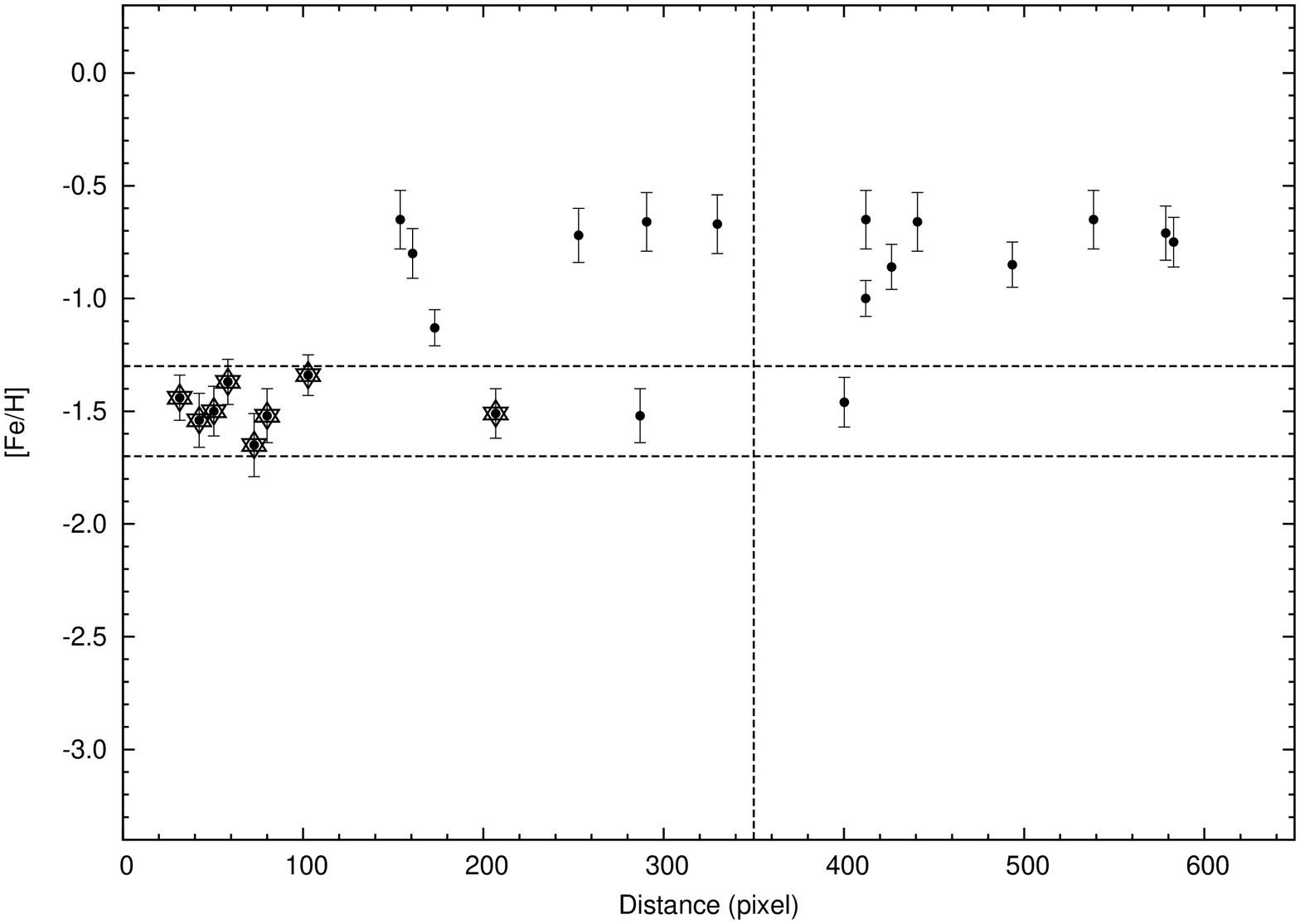}
\includegraphics[height=8cm,width=6cm,angle=-90]{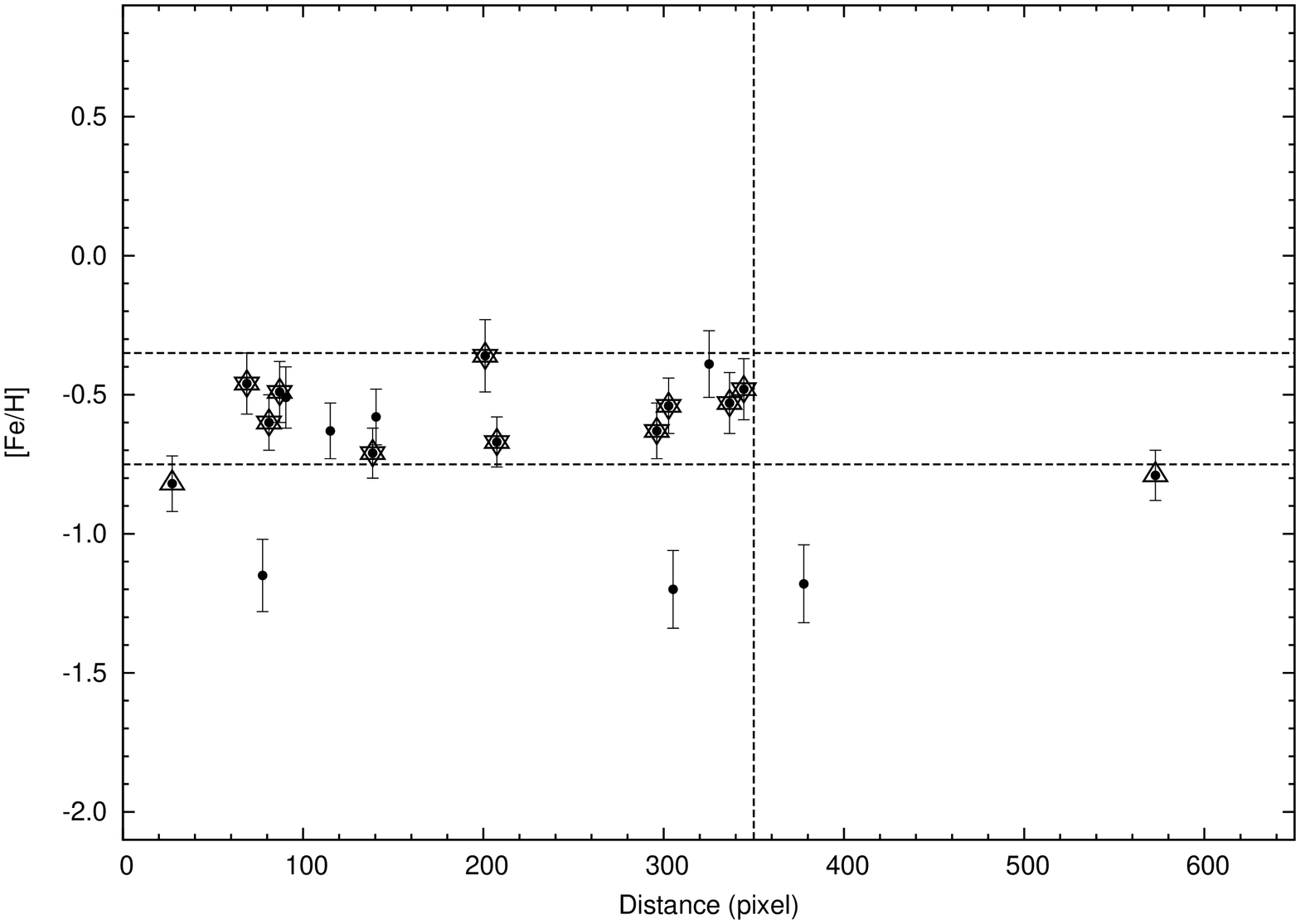}
}
\caption{Metallicities as a function of distance for all target stars in NGC 1754 ({\it left}) and NGC 1900 ({\it right}).
The [Fe/H] cut of $\pm0.20$ dex is denoted by the horizontal lines. For NGC 1754,
a metal-poor cluster, the field  is relatively easily distinguished from the
cluster. The plotted error bars represent the systematic error in [Fe/H] calculation. 
Triangles are stars  satisfying the radial velocity criteria 
and the star symbols are those which qualify both selection criteria based on radial velocity and [Fe/H].
}
\label{eq}
\end{figure*}

Cluster membership criteria are very crucial in defining the parameters for the clusters.
We use a combination of several criteria to isolate cluster members 
from field stars as defined by G06.  
First, all observed targets were selected on the base of the statistically decontaminated near-IR
CMDs (Kurtev et al., 2010). Second, the cluster centers 
and radii are chosen by eye, based primarily on the photometric catalog ($VR$). 
As an example, Figure \ref{map} {\it(left panel)} shows (x, y)-positions
for all stars photometered in the NGC 1754 and NGC 1900 fields respectively, with large dots
denoting our target stars and the large open circle representing 
the adopted cluster radius; stars outside this circle
are considered non-members due to their large distance from the cluster center. 
We note that stars outside the cluster radius were observed so as to define parameters for the
LMC field, which aids in isolating cluster members. 
Next, radial velocity versus radial distance is plotted in Figure \ref{map} {\it(right panel)}. Stars moving with
the velocity of the clusters are easily identified due to their smaller
velocity dispersion. Our velocity cut, denoted by the horizontal lines, has been
chosen to include   the expected observed velocity dispersion in
each cluster. To determine this, we have adopted an intrinsic cluster velocity 
dispersion of 5 km $s^{-1}$ and added this in quadrature
with our adopted radial velocity error, 9 km $s^{-1}$, resulting in
a dispersion of $\sim10.3$ km $s^{-1}$. Thus, 
we adopted a width of $\pm$11  km $s^{-1}$ for our radial velocity cut.
We select the cluster velocity to maximize the number of stars lying within this range.
The cluster radius (Figure \ref{map} {\it(right panel)}, vertical line) is also marked for reference.
Finally, Figure \ref{eq} shows metallicity as a function of distance for the
stars in NGC 1754 and NGC 1900 respectively, with horizontal lines representing the [Fe/H]
cut of $\pm0.2$ dex that has been applied to these data.
We adopt random error of $\pm 0.15$ dex in measuring each individual [Fe/H],
this along with a typical systematic error of $\pm0.11$ dex gives
total error in measurement $\pm0.18$ dex. Thus, we have rounded this figure to $\pm0.2$ dex.
Radial velocity members which are in this range of metallicity are assumed to be probable cluster member.
Star symbols denote stars that have made all of these cuts and are therefore considered to be cluster members. 

\begin{figure*}
\centering
\includegraphics[height=6cm,width=8cm]{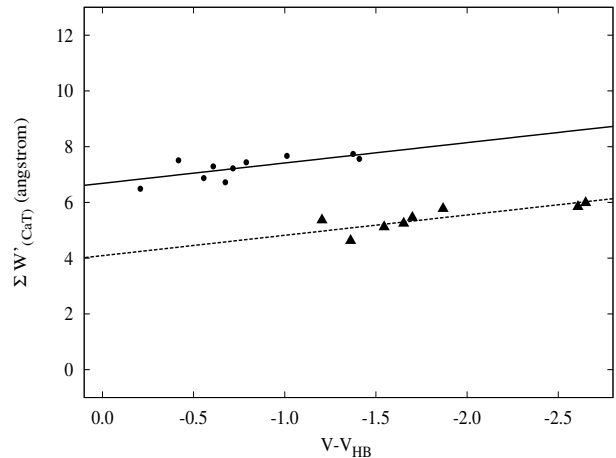}
\caption{ Summed equivalent width vs. brightness above the HB for all stars
considered to be members of NGC 1900 (circle) and NGC 1754 (triangle). 
The slanted lines are iso-abundance line with a slope $\beta$ = 0.73 at
the mean metallicity of the clusters i.e. [Fe/H] = $-0.55$ and $-1.48$ for NGC 1900 and NGC 1754 respectively.  }
\label{hb}
\end{figure*}

\begin{figure*}
\centering
\hbox{
\includegraphics[height=8cm,width=8cm,angle=-90]{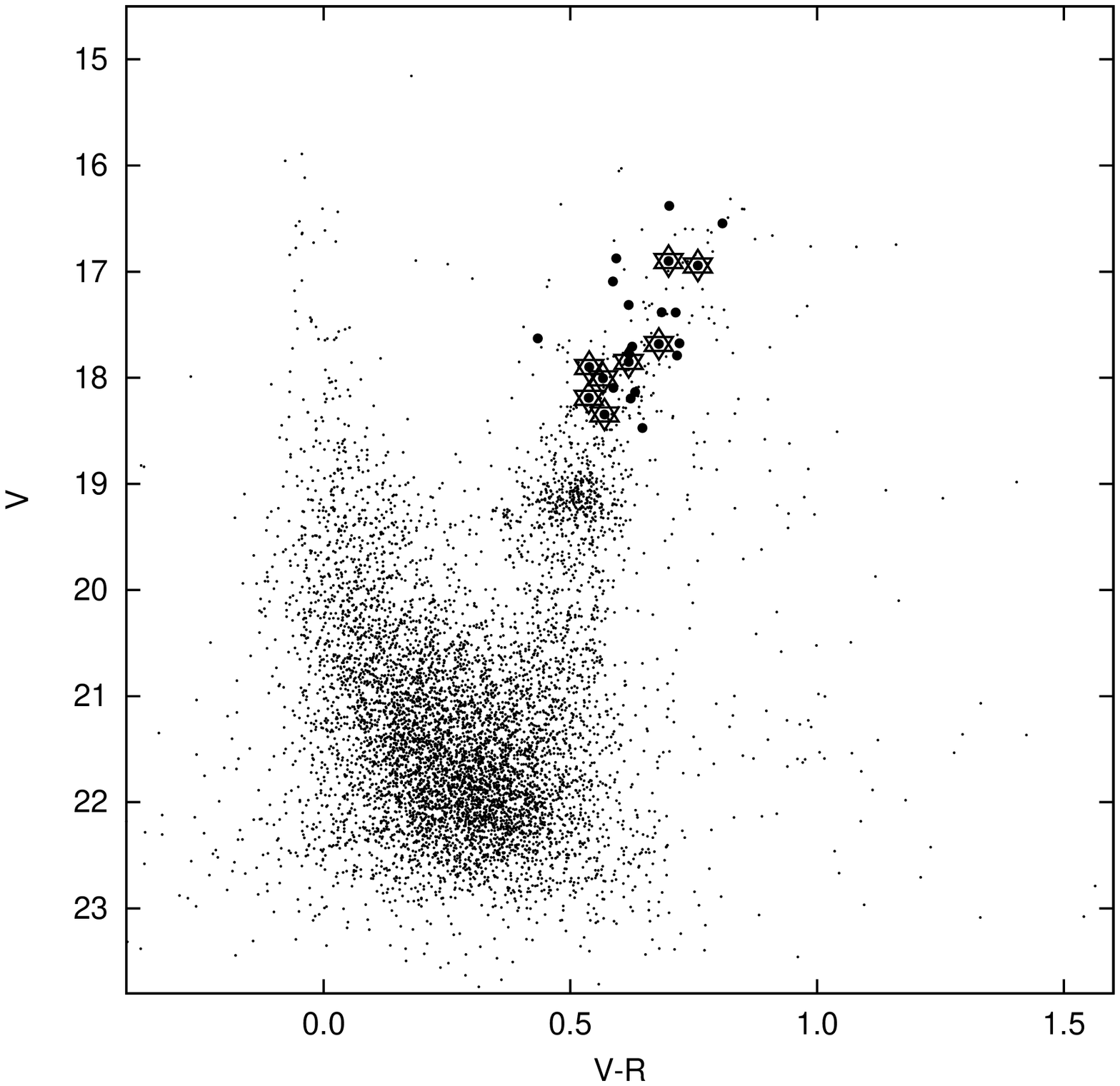}
\includegraphics[height=8cm,width=8cm,angle=-90]{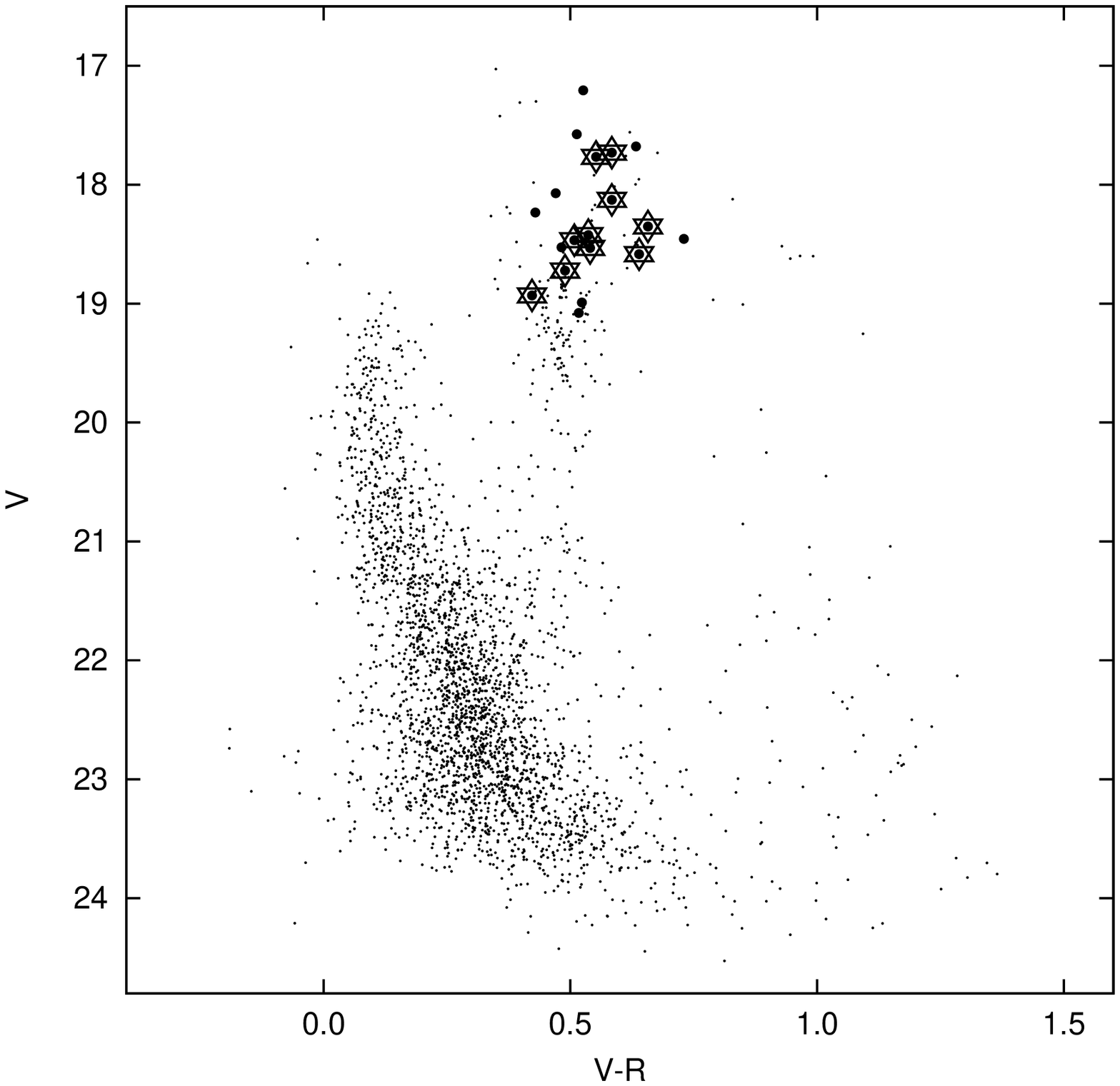}
}
\caption{CMD for the entire NGC 1754 ({\it left}) and NGC 1900 ({\it right}) field, with CaT stars marked with big dots.
Confirmed cluster member RGB stars are also shown by star symbols.
}
\label{cm}
\end{figure*}

In Figure \ref{hb} we present the traditional $\Sigma$W versus $V-V_{HB}$ plot for
cluster members of NGC 1754 and 1900 with the lines representing the mean
metallicity of $-1.48$ and $-0.55$, respectively, with a standard error of the mean =0.03. The CMD in Figure \ref{cm} shows all stars
photometered along with CaT targets (big dots) in the NGC 1754 and NGC 1900 field respectively.
Confirmed  member RGB stars are also shown by star symbols.

In Table \ref{data}, for all RGB stars of the observed LMC 
clusters, we list the following information: stellar
identification number, RA, Dec, heliocentric radial velocity and its  
associated error, $ V - V_{HB}$, S/N, $\Sigma$W, W$^\prime$ along with errors,
and the value of [Fe/H] with associated error.

\section {Cluster Properties}

\subsection{Color-Magnitude Diagrams}

\begin{figure*}
\centering
\includegraphics[height=15cm,width=15cm,angle=-0]{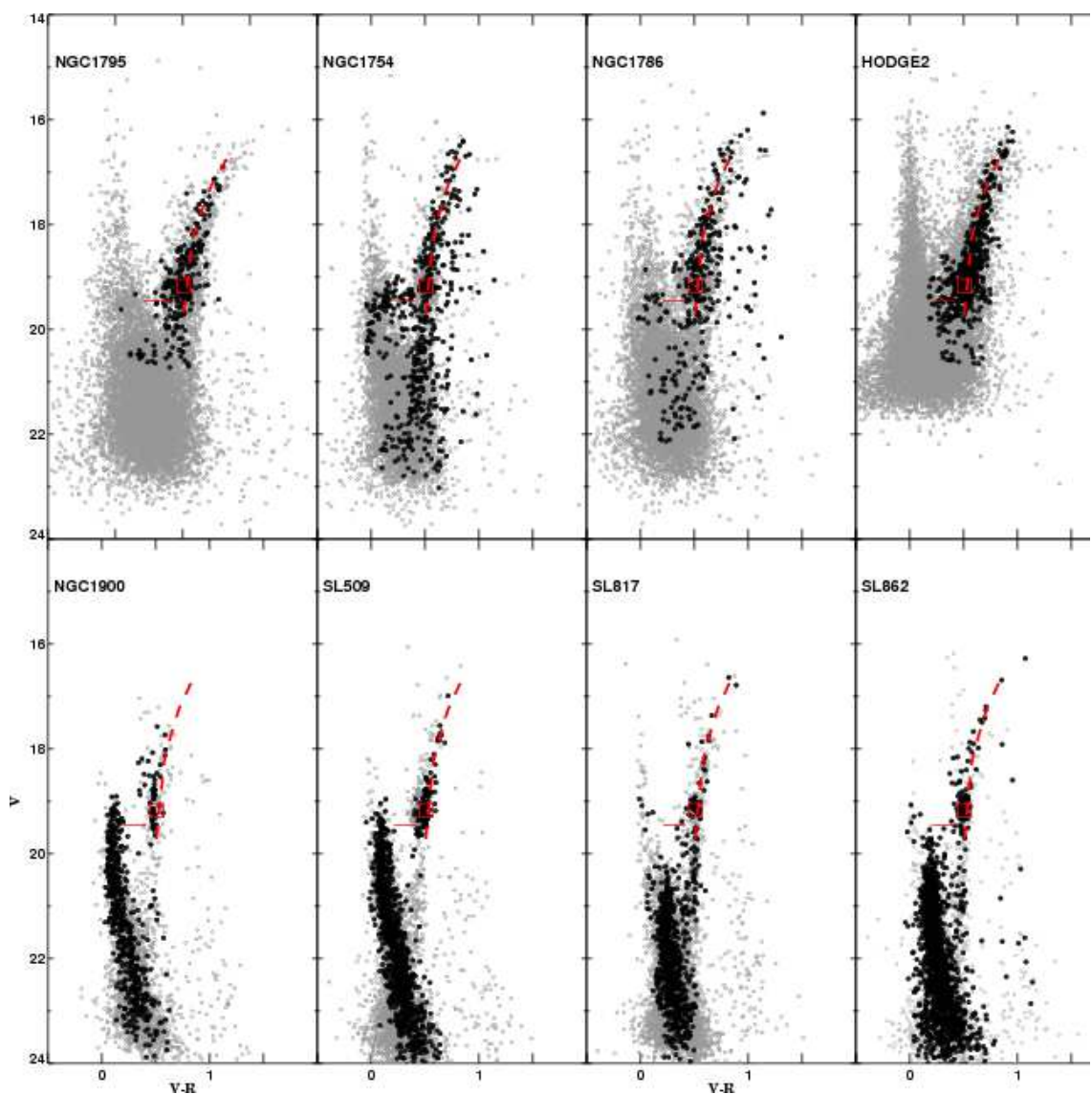}
\caption{CMDs of NGC\,1795; NGC\,1754; NGC\,1786; Hodge\,2; NGC\,1900; SL\,509; SL\,817; and SL\,862. The grey dots represent all stars  from our photometric lists, the 
most probable cluster members (see the text) are  plotted with dark dots. The dashed and solid lines show the MACHO mean RGB region and the location of the field RR Lyrae star, the big square  shows MACHO RC region.}
\label{cmd_all}
\end{figure*}

Figure~\ref{cmd_all} shows the CMDs of the eight LMC clusters. All the stars 
measured with good precision (error in mag $<$0.1) in the whole FORS2 frames are plotted here: therefore the CMDs 
include both cluster and field stars. To check the zero point of our calibrations we  over-plotted the fiducial 
lines of RGB stars, field RR Lyrae and Red Clump (RC) stars as defined from MACHO database (Alcock et al. 2000, see their 
Table~1.). 

The clusters from our sample are located in areas densely populated by field
stars. In order to obtain a cleaner cluster CMD, we proceeded in the following way.
For each cluster, an annulus around the center was selected with the
criterion of being large enough to contain most cluster stars, but
small enough to minimize the background contribution. The small circular region
containing the very center of some of the clusters was excluded because 
these areas were too crowded, and therefore the photometry was
rather poor. The photometry around very bright stars projected close to the objects NGC\,1786, SL\,826, and NGC\,1754 is also removed. A CMD typical of LMC field stars was extracted from a second region, well outside a given radius around the cluster center,
selected with the criterion of containing a number of stars large
enough to insure good statistics, but far enough from the cluster not
to contain a significant fraction of cluster stars. The cluster and reference fields
cover the same area. The full range of magnitude and colors of a given CMD was divided into grids and we
computed the expected number-density of field stars in each cell based on the number of 
comparison field stars. After that the expected number of field stars is randomly  subtracted 
from each cell. The statistically de-contaminated CMDs are plotted in
Figure~\ref{cmd_all} as dark dots.    

\begin{figure*}[h]
\includegraphics[height=8cm,width=8cm,angle=0]{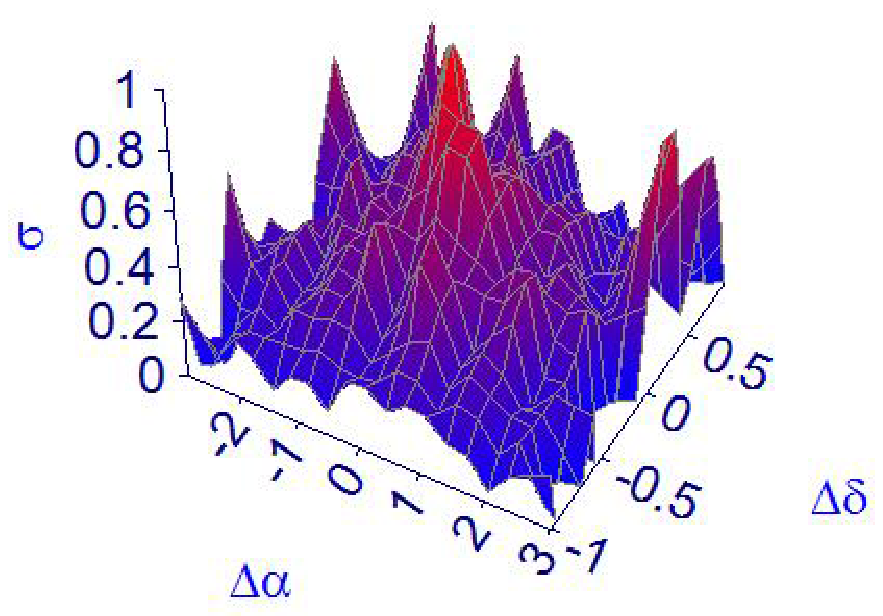}
\includegraphics[height=8cm,width=8cm,angle=0]{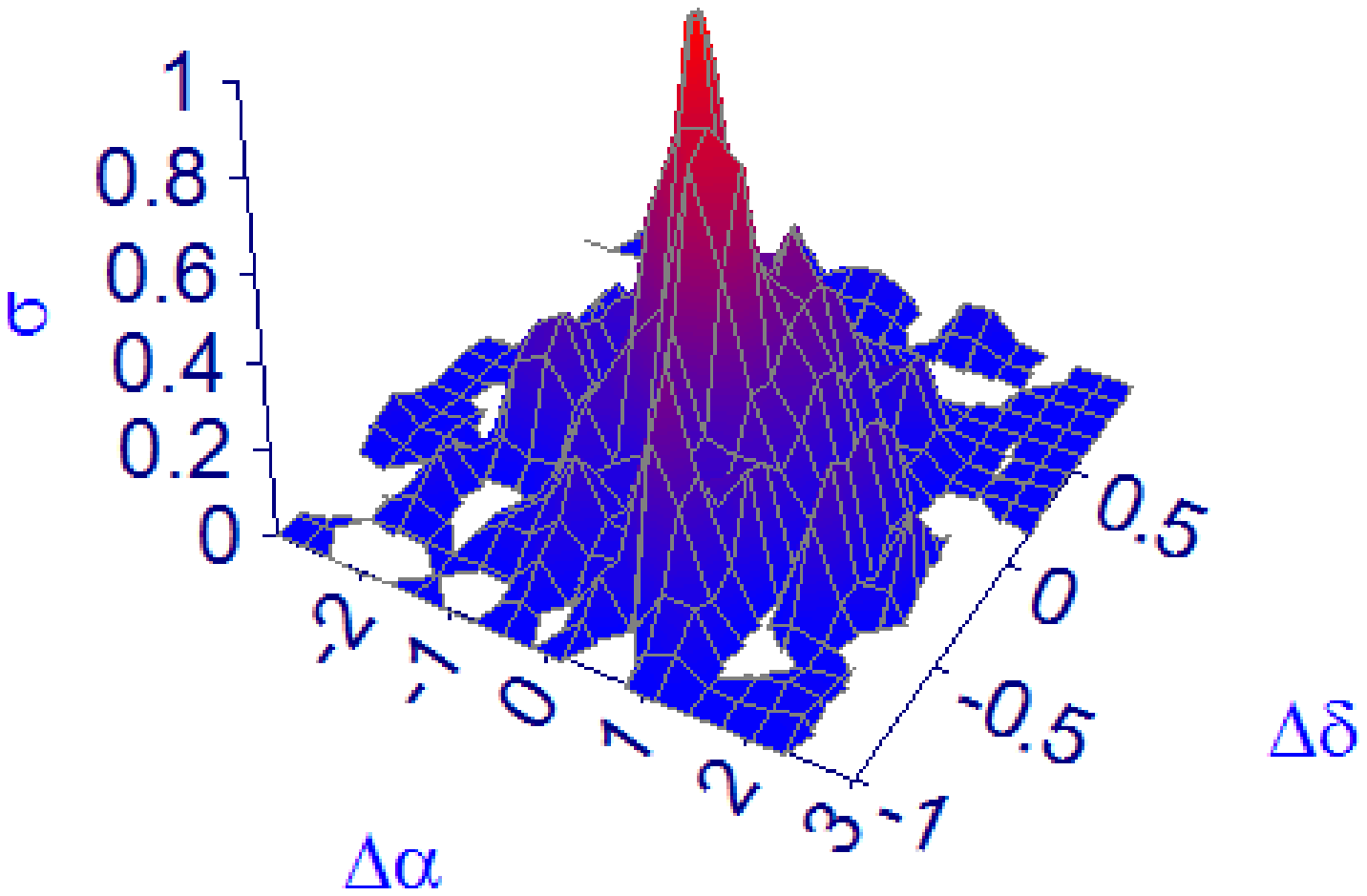}
\caption{The stellar surface-density distributions (number of stars per arcmin$^2$) of NGC\,1795. {\it Left panel}: All stars in our photometric list. {\it Right panel}: 
Most probable cluster members.}
\label{radec_count}
\end{figure*}

To check our de-contamination procedure we calculated the stellar surface-density 
distributions (number of stars per arcmin$^2$) (Bonatto \& Bica, 2006). 
They are computed for a mesh size of $0.2 \times 0.2$  arcmin$^2$, centered on the apparent cluster center.
Figure\,\ref{radec_count}, left panel shows observed photometry of NGC\,1795, while on the right-hand panel only the stars selected by means of the color-magnitude filter described above are plotted. As can be seen, most of the field stars have been removed, although a small number of field stars still fall in the cluster main sequence locus region.

The decontaminated CMDs outline the basic features of the clusters. Most of them show typical diagrams of intermediate age clusters. As can be seen in Figure~\ref{cmd_all_cl} we can not reach the Turn off/Termination  Point (TO)
for NGC\,1795; NGC\,1754; NGC\,1786; Hodge\,2 because of the crowding and short exposure times. 
For the rest of the clusters from our sample: NGC\,1900; SL\,509; SL\,817; and SL\,862 the main sequence 
can be followed up to 4 magnitudes bellow the TO. The RC stars are located between $\approx$ 19-19.3 mag and 
the RGB extends between 2 and 4 magnitudes brightward of the RC. NGC\,1754, and to a lesser extent NGC\,1786, show well populated horizontal branches (HB).

\subsection{Age determination}

The goal of this project is to obtain homogeneous [Fe/H] values and ages to be used for calibration of the near-infrared $K_s$-band RGB slope-metallicity-age relation. The determination of the metallicities is described in \S 4.
The most reliable technique for age determination is to examine the luminosity of the main sequence turnoff and compare it to the predictions of theoretical isochrones. Unfortunately, it is not possible for half of our objects, since
the photometry is just not deep enough and we could not reach the TO and main sequence region. We choose to compare the rest of our CMD with the Padova set of theoretical isochrones (Marigo et al. 2008), because they are based on models with a moderate amount of convective overshooting, asymptotic giant branch evolution and circumstellar dust around the coolest evolved stars. For the clusters NGC\,1795, NGC\,1754 and NGC\,1786 much deeper, better HST data are available in the literature and we adopted the age determination from the corresponding papers. Since in these papers different sets of theoretical isochrones are used, we plotted the corresponding isochrone from the Padova set to check the coincidence. 
To determine the best models we applied a statistical tool to compare the observed and theoretical main sequence and/or  red clump. We compared the differences along the observed and theoretical fiducial lines of each cluster
using a $\xi^{2}$ statistics.  

\begin{figure*}
\includegraphics[height=15cm,width=15cm,angle=0]{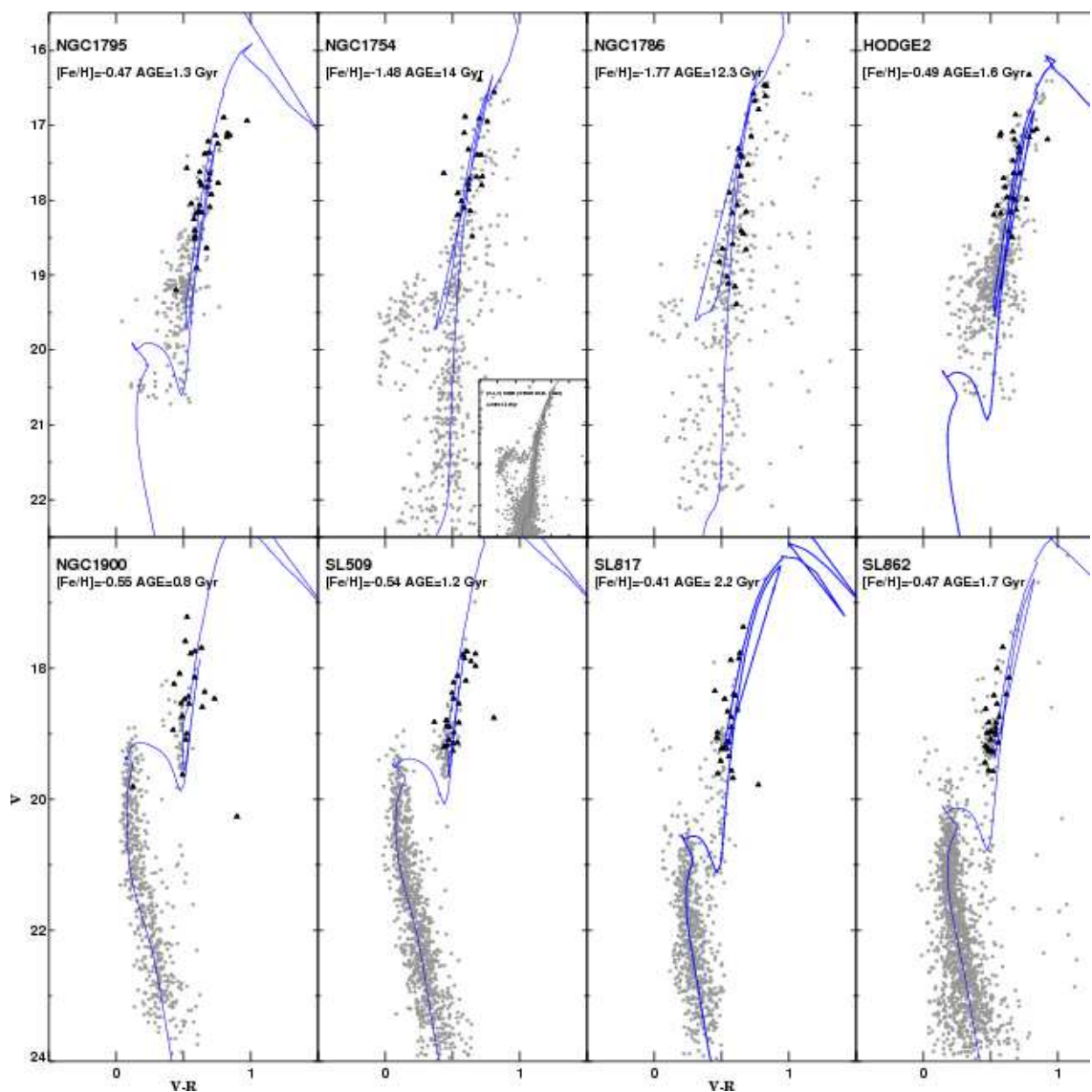}
\caption{The statistically decontaminated CMDs of NGC\,1795; NGC\,1754; NGC\,1786; Hodge\,2; NGC\,1900; SL\,509; SL\,817; and SL\,862. The dark triangles represent 
spectroscopically observed stars. The solid lines show best fit of the Padova set of theoretical isochrones (Marigo et al. 2008), whose age and metallicity are marked.
The smaller panel for NGC 1754 represents the HST $VI$ data for the cluster.
}
\label{cmd_all_cl}
\end{figure*}

{\it NGC\,1795}: 
Because of the crowding, incompleteness and short exposure times of our images the CMD
of NGC\,1795 does not reach the TO. The RGB and RC stars however are clearly visible and the corresponding branches are well populated. To determine the RC brightness we adopted the  method described in Pietrzy{\'n}ski et al. (2003): we selected the stars 
with $V$-band magnitudes and $V-R$ colors in the range around the  RC stars and fit the derived  $V$-band magnitude histogram with a complex function involving a Gaussian function (representing the distribution of RC stars) 
superimposed on a second-order polynomial component (for the stellar background; 
see, e.g., Stanek \& Garnavich 1998; Alves et al. 2002; Pietrzynski \& Gieren 2002).
Such a described procedure finds  the RC at $V$=19.13$\pm$0.07.
The latest photometric investigation based on the archival HST images of the cluster is given in Milone et al. (2009). 
The observed deep ($V,~ V-I$) CMD is compared with a set of BaSTI isochrones and gives the following 
parameters: $(M-m)_0$=18.45; $E(B-V)$=0.10, z=0.008 and age 1.3 Gyr. Since we have accurately determined 
metallicity [Fe/H]=-0.47 we adopted their distance, reddening and age values and plot the corresponding 
isochrone from Padova set which fit well the luminosity and color of the RC region (Figure~\ref{cmd_all_cl}, upper left panel).  

{\it NGC\,1754}:
NGC\,1754 is well known old LMC cluster (Olsen et al. 1998). The main features visible on our CMD are well populated RGB, long blue HB, typical for 
metal poor clusters and a sub-giant branch (SGB) down to $V$=22 mag. The TO point and RC regions are not visible. In our photometry, the RC is determined at $V$=19.25$\pm$0.17. The level of the HB calculated as lower envelope is found at $V$=19.55$\pm$0.11. Olsen et al. (1998) derived $(M-m)_0$=19.00 (based on the MW globular cluster stars comparison)  and $(M-m)_0$=18.62 (based on the level of the HB stars); $E(B-V)$=0.10 and age 15.6 Gyr. We adopted their average distance value of $(M-m)_0$=18.66 and plotted the Padova's isochrone of 15.6 Gyr, but the deviation from the observed CMD is very large. Taking into account our metallicity [Fe/H]=$-1.48$ the best fit of the Padova isochrones (Marigo et al. 2008) favorite an age of 14 Gyr. To check this age determination we retrieve the HST $V,I$ photometry of Olsen et al. (1998). As can be seen in Figure~\ref{cmd_all_cl} (right corner of NGC\,1754 plot) an age of 14 Gyr fits well the TO region of the Olsen et al. (1998) CMD. Obviously, this is slightly older than  the well-determined age of the Universe from WMAP and therefore an overestimate.

{\it NGC\,1786}: 
This object is another example of an old, metal poor cluster in the LMC (Brocato el al. 1996). In our CMD the RGB is well defined, the HB is populated on the blue side of the instability trip at  $V$=19.44$\pm$0.08 (lower envelop). The fundamental properties of NGC\,1786 are taken from Table~7 of Grocholski et al. (2007) and are based on the mean apparent $V$-band magnitude of RR Lyrae stars (Walker 1985; Walker \& Mack 1988; Walker 1989, 1990, 1992ab, 1993).
With  $(M-m)_0$=18.62; $E(B-V)$=0.06 and [Fe/H]=-1.77, we plotted Brocato el al. (1996) isochrone of 12.3 Gyr. As can be seen in Figure~\ref{cmd_all_cl} the isochrone for the adopted age fits well our CMD.  
 
{\it Hodge\,2}: 
There is no photometric information for this cluster in the literature.
Our CMD does not clearly reach the TO point; however, the RGB is well defined
and  the RC can be located at
$V_{RC}$= 19.12$\pm$0.17.
To determine the age of the cluster we need reddening and distance. We calculate the distance using the Population II Cepheid star SV*HV 2439 (OGLE LMC-SC7 21841), a confirmed cluster member (Hodge \& Wright 1963; Udalski et al. 1999). According to OGLE III database (Soszynski et al. 2008) the apparent mean $V$ magnitude of the star is $V$=15.58, with a period of pulsation of P=4.81308 days.  The reddening  $E(B-V)$=0.14 is adopted from the OGLE database as a mean value of the SC7 field. The distance modulus $(M-m)_0$=18.62 is calculated as a difference between reddening corrected mean $V$ magnitude of SV*HV 2439 and the absolute $M_V$ magnitude of Cepheid stars given in Benedict et al. (2002). 
The cluster has a metallicity value [Fe/H]=$-0.49\pm0.11$ and the best fit 
isochrone favors an  age of 1.6 Gyr. 
 
{\it NGC\,1900}: 
The CMD presented here is the only photometric investigation up to the moment and shows typical features of an intermediate age stellar cluster: long and well populated main sequence reaching about 4 magnitudes below the turnoff     point and some evolved RGB stars. The distance of $(M-m)_0$=18.47$\pm$0.07 is calculated from the mean RC $K_S$ brightness derived from our high quality $J,K_S$ infrared photometry (Kurtev et al. 2010). We used the absolute RC magnitude $M_K = -1.644$, calibrated in  Alves et al. (2002). To calculate the distance a correction for population effects has to be applied (Gullieuszik et al. 2007), because of the differences in the stellar content of LMC and Galactic RC stars on which the Alves (2002) calibration is based. Salaris \& Girardi (2002) estimated the population effects on the RC absolute  magnitude in the $K$ band of $\Delta M_{K}^{RC}=-0.03$. Using [Fe/H]=$-0.55\pm0.09$ obtained in \S 5, we derive an age of 800 Myr.

\begin{figure*}
\includegraphics[height=7cm,width=15cm,angle=0]{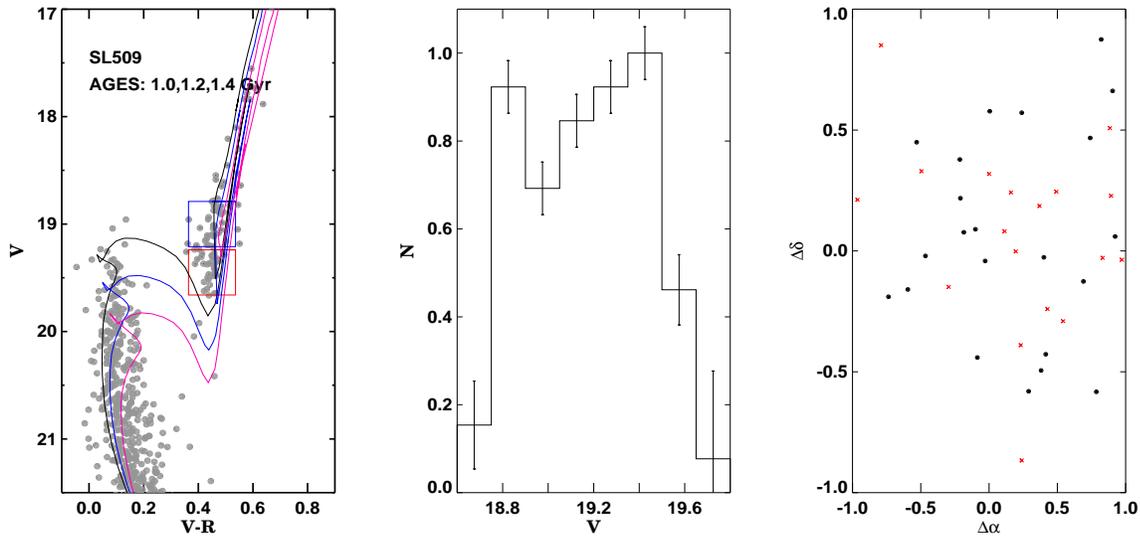}
\caption{ {\it Left panel}: CMD of SL\,509. The regions of the two suspected RCs are marked with large squares. The solid lines show the Padova theoretical isochrones 
(Marigo et al. 2008) for ages of 1.0, 1.2 and 1.4 Gyr. {\it Middle panel}: The luminosity function of the RC region. {\it Right panel}: 
The positions of the main RC stars (filled circles) and suspected secondary clump (crosses).
}
\label{sl509_cmd}
\end{figure*}

{\it SL\,509}: 
The only photometric investigation up to the moment is reported by Bica et al. (1998) and Piatti et al. (2003) and is based on the Washington photometric system. The derived age in Piatti et al. (2003) is 1.2 Gyr and the metallicity
corrected for age effects is [Fe/H]=$-0.65$. They also reported a secondary clump 0.45 mag fainter and bluer than the main RC, which also can be noted in our CMD. In Figure~\ref{sl509_cmd} (left panel) we zoomed the RC and TO regions of the CMD, marking with boxes the suspected clumps. The luminosity function of the RC region is shown in the middle panel of Figure~\ref{sl509_cmd} and finally the last panel of the Figure~\ref{sl509_cmd} represents the positions of the stars with respect to the cluster center. As can be seen there is a strong hint of two populations, as pointed out by Piatti et al. (2003), but the sample is too small with only 66 stars and more data are necessary  to confirm the reality of these populations. If confirmed, according to Girardi et al. (2009) the secondary clump could be a result of He-ignition in stars just massive enough to avoid $e^-$-degeneracy in their H-exhausted cores. 
The TO can be located at $V_{TO}$= 19.40$\pm$0.11. Using the same method as for NGC\,1900 we derived $(M-m)_0$=18.48$\pm$0.05. The E(B-V)=0.03 is taken from the OGLE database. The best isochrone fit gives 1.2 Gyr, in excellent agreement with Piatti et al. (2003) result.

{\it SL\,817}: 
The cluster has only Washington photometry (Bica et al. 1998; Piatti et al. 2003) and shows morphology similar to that of 
SL\,509. Using the same technique as for SL\,509 we derived the TO at $V_{TO}$= 20.20$\pm$0.11, the RC is calculated at $V_{RC}$= 18.96$\pm$0.17. For the distance, reddening and age we determined $(M-m)_0$=18.38$\pm$0.07; E(B-V)=0.12 and 2.2 Gyr, respectively.
 
{\it SL\,862}: 
And finally for SL\,862 we have TO at $V_{TO}$= 20.10$\pm$0.11; the RC  at $V_{RC}$= 19.07$\pm$0.09; $(M-m)_0$=18.45$\pm$0.07; E(B-V)=0.10 and an age of 1.7 Gyr. Piatti et al. (2003)
give an age of 1.7 Gyr and the metallicity of $-0.75$ from Washington photometry, in agreement with our values.
 
We estimated the errors in the age determination as an upper limit of the various factors: errors in the photometry, errors in the transformation to the standard system, errors due to incompleteness, field decontamination, metallicity determination and errors of the fit to the theoretical isochrones. For the clusters without 
good main sequence photometry, ages can be estimated to
within $\pm$ 1.4-1.7 Gyr, while for the younger clusters
with prominent main sequences we estimate  an error of $\pm$0.5 Gyr.
In  Table \ref{param}, we have listed the obtained value of age for all the target clusters. 
The radial velocity and metallicity along with standard deviation is also given in Table \ref{param},
which has been obtained by taking the mean of the individual values of cluster members.

\section {Analysis}

\subsection{Comparison to previous metallicity determination and sample enlargement}

We compared the values of the metallicities obtained in the present work with previously published values.
Since our method relies on high S/N spectroscopy of a large number of stars per cluster,
the individual and mean errors are in  general smaller as compared to the previously determined values
(for eg. O91 have used in general 2 stars per cluster, while we have on average 10 identified cluster members).
Four of our target clusters have metallicity values given in O91 using CaT spectroscopy of some individual giants.
Five of our target clusters also have photometrically determined values of metallicity (see Table \ref{param}).
As indicated by G06, although O91 used CaT lines as a proxy for measuring Fe abundance
directly and is thus similar to our work, it is inappropriate to directly compare the O91 values to our work due to differences in the details of the 
metallicity determination (see for detail G06).
C05 find that one can estimate the abundance of O91 clusters on the metallicity system we have used 
via the following conversion:

[Fe/H] $\simeq$ - 0.212 + 0.498[Fe/H]$_{\rm O91}$ - 0.128[Fe/H]$^2_{\rm O91}$

\begin{figure*}
\centering
\includegraphics[height=6cm,width=8cm]{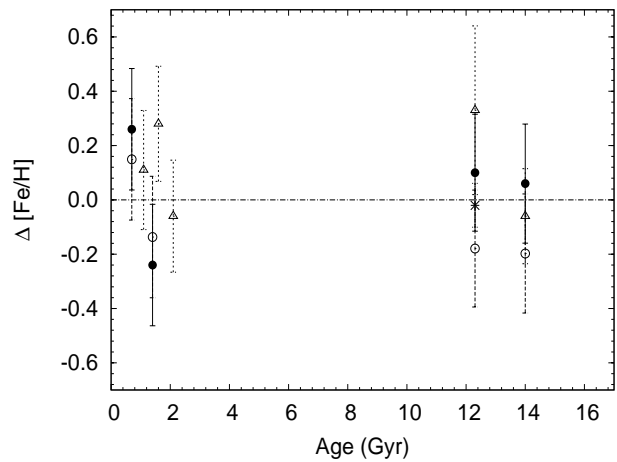}
\caption{Difference between present and previously determined [Fe/H] value ($\Delta$ Fe/H = Fe/H$_{present}$ - Fe/H$_{literature}$) vs. Age diagram.
Filled and open circles represent original and transformed O91 values respectively and
open triangles are for photometrically observed clusters.  Cross marks the
high resolution value given for NGC 1786 by Mucciarelli et al. (2009). } 
\label{cmp}
\end{figure*}

In Table \ref{param}, column (6), (7), (9) gives the [Fe/H] values obtained in 
the present study,
values given by O91 (spectroscopically) and the photometrically determined values, respectively.
In Column (8), we give the transformed values of [Fe/H] given by O91 to our system using the above relation.
In Fig. \ref{cmp}, we have plotted the difference `${\rm [Fe/H]} _{present}~-~{\rm [Fe/H]}_{literature}$' as a 
function of  age for O91 clusters (filled circles: original O91 values, open circles: 
transformed O91 values) and photometrically observed clusters (open triangles). 

The difference of metallicity between the present value and  the transformed value of O91
are of the order of $0.15-0.23$ dex. O91 give their [Fe/H] errors for individual stars as 0.2 dex, therefore,
deviations between these data sets as large as $\pm0.2$ are not unexpected, suggesting that these results are in 
relative agreement with no offset. We note, however, that even with the use of the above equation, it is very difficult to directly 
compare the derived cluster abundances because of the difference in calibration strategy.
Comparison of our [Fe/H] values with the photometrically determined values for each cluster is given below:

NGC 1754: This cluster has  been studied by Olsen et al. (1998) using HST CMDs quoting a metallicity of $-1.42\pm0.15$.
Olsen et al. (1998) commented that the abundances they obtained are on average higher than previously measured spectroscopic abundances.
They have used the method of Sarajedini (1994) to determine abundances from the measurement of the height of the RGB above the level of HB.

NGC 1786: This cluster has been studied by Brocato et al. (1996)
using $VI$ data from the 3.6 ESO NTT. They have determined a lower value of $-2.1$ but with a large error (0.3 dex) using the Sarajedini (1994) method. 
Recently this cluster was also studied by Mucciarelli et al. (2009) using 
high-resolution spectroscopy
of giant stars. Their value of [Fe/H]= $-1.75\pm0.01$ is virtually identical
to our metallicity determination ([Fe/H] =$-1.77\pm0.08$), lending confidence 
to our results.

SL509, SL817 \& SL862: These clusters have been studied by Piatti et al. (2003) using Washington photometry.
Our  [Fe/H] values for SL509 and SL817 (i.e. $-0.54\pm0.09$, $-0.41\pm0.05$) are comparable to their reported 
values of  $-0.65\pm0.2$ and $-0.35\pm0.2$, respectively.
For SL862, they gave [Fe/H] = $-0.75\pm0.2$, which is lower than the present estimation of $-0.47\pm0.07$.

For all eight clusters, we have used the very reliable CaT method as the [Fe/H] 
indicator,
with significantly more cluster members and  less scatter in individual [Fe/H] value compared to previous studies.

It is important to have significant numbers of data points to make a definitive
analysis of the age-metallicity relation of LMC clusters. 
Over the past few decades, nearly a hundred LMC star clusters have had both accurate age estimated from 
isochrone fitting to their CMDs (Sagar \& Pandey 1989;  Balbinot et al. 2009; and references therein),
and a metallicity measurement from either spectroscopy and/or
isochrone fitting (G06).  Recently Harris \& Zaritsky (2009) have compiled the age and metallicity of 85 clusters
from their literature search. Out of these 85, 5 are in common with our sample and 15 are in common with G06.
We have excluded our 5 and  15 G06 clusters from this sample and generate a new sample of 101 clusters
[(85 -5 -15) + 8 + 28 = 101]. The present work combined        with G06 makes a homogeneous dataset of  36 clusters
using the same method for metallicity determination.
In the case of G06 clusters, out of 28, 15 have ages from the sample of Harris \& Zaritsky (2009). For the other 13
clusters, we converted the given SWB type (Searle, Wilkinson, \& Bagnuolo 1980, hereafter SWB) into age from the conversion given in Bica et al. (1996).
These 13 clusters have either SWB type V or VI. Based on published data, we have taken the mean age 
of SWB type V and VI as 1.5 and 2.5 Gyr respectively. 
Here it is important to mention that the Harris \& Zaritsky (2009) compilation is heterogeneous, using 
a variety of different age and metallicity scales, so in fact it is a bit dangerous
lumping them all together, although this effect is minimal for the present study.

\subsection{ Age/Metallicity Distribution and their  Relationship}

The cluster metallicity distribution (hereafter MD) is an important diagnostic of the
global chemical evolution of a galaxy and is useful for an overall comparison of the cluster
systems of different galaxies (P09).

\begin{figure*}
\centering
\includegraphics[height=8cm,width=6cm,angle=-90]{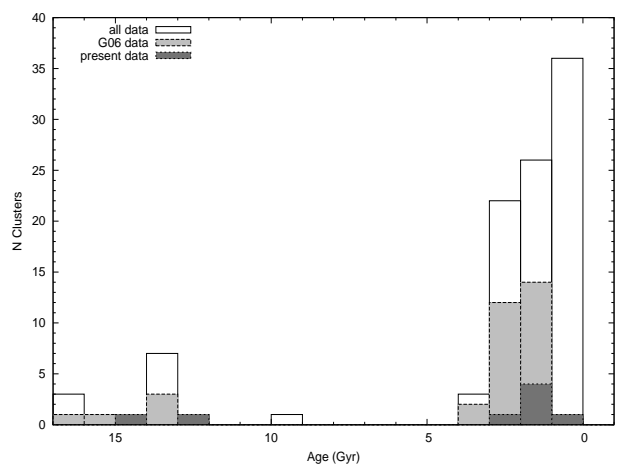}
\caption{Age distribution of LMC clusters.}
\label{hist_age}
\end{figure*}

\begin{figure*}
\centering
\includegraphics[height=8cm,width=6cm,angle=-90]{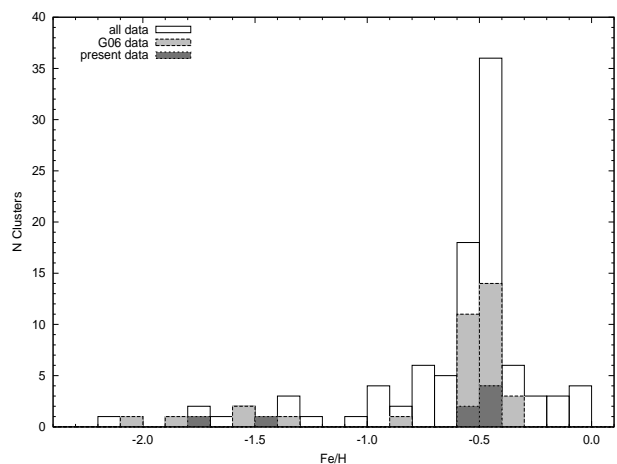}
\caption{Metallicity distribution of LMC clusters.}
\label{hist}
\end{figure*}

In Figure \ref{hist_age} we have shown first the distribution of the ages of all 101 clusters.
The black, grey and open regions represent the data from the present study, from G06 and from Harris \& Zaritski (2009), respectively.
Broadly we can classify the population in three age bins: $<1$, $1-3$ and $>12$ Gyr,
ie. youngest, intermediate age and old age populations.
In Table \ref{stat}, we have presented the statistics for the value of [Fe/H] and age in these
bins along with the number of clusters using the CaT  method or clusters from the
literature survey. We can easily see the change in the value of [Fe/H] with age.
Globally, for clusters having mean age $\sim 0.2$ Gyr, the [Fe/H] value is $-0.35$,
for age $\sim 2$ Gyr, the [Fe/H] value is $-0.55$ and for age $\sim 14$ Gyr, the [Fe/H] value is $-1.62$.
This change is more prominent in $2^{nd}$ and $3^{rd}$ bin of clusters having used the CaT method 
(we have only one cluster in $1^{st}$ bin). 
We found that the younger clusters are generally more enriched than older clusters with the
youngest bin [Fe/H] value comparable to the present-day LMC    value of `$-0.31\pm0.04$' given by Rolleston, Trundle \& Dufton (2002) and 
 `$-0.27\pm0.06$' given by Hill et al. (1995).
The MD is shown in Figure \ref{hist}, where the 
color code is the same as Figure  \ref{hist_age}.
We can easily see the strong peak in the distribution of clusters having [Fe/H] = $-0.5$ to $-0.4$ and 
rest of them are distributed approximately uniformly with metallicity. 
The metallicity of the six young clusters obtained in the present study, having 
a mean age of 1.5 Gyr, is $-0.49$ with a scatter of
0.04, in excellent agreement with the previous finding of G06, which shows a very tight distribution
with mean metallicity of $-0.49$ ($\sigma$ = 0.09) for the younger clusters, which formed after the age gap ended some 3 Gyr ago.
The narrow metallicity distribution of LMC young-intermediate age clusters has also been observed previously,
based on high resolution spectroscopy (i.e. NGC 1866: [Fe/H] = $-0.5\pm0.1$ (Hill et al. 2000) and 
mean [Fe/H] = $-0.38\pm0.09$ for 4 LMC globular clusters (Mucciarelli et al. 2008)).
This agrees well with the theoretical work of Bekki et al. (2004), which indicates that a
close encounter between the LMC and SMC caused not only the restart of cluster formation in the
LMC but the generation of the central bar (mean [Fe/H] = $-0.37\pm0.15$; C05).
We can easily see the two separate distributions in the age of the clusters (Figure \ref{hist_age}) and
older clusters are showing a much larger spread in [Fe/H] value as compared to younger ones 
(see also  Figure \ref{hist}). 
The sample size is near the limit of drawing statistically
significant results of this kind. 
It is not clear why the SMC cluster age and metallicity distribution are distinct
from that of the LMC.
It is of great interest to enlarge the present homogeneous sample in order to 
ascertain the LMC cluster MD.
Note that the CaT data shows a much tighter metallicity distribution than that
of Harris \& Zaritsky (2009) or O91.

\begin{figure*}
\centering
\includegraphics[height=8cm,width=6cm,angle=-90]{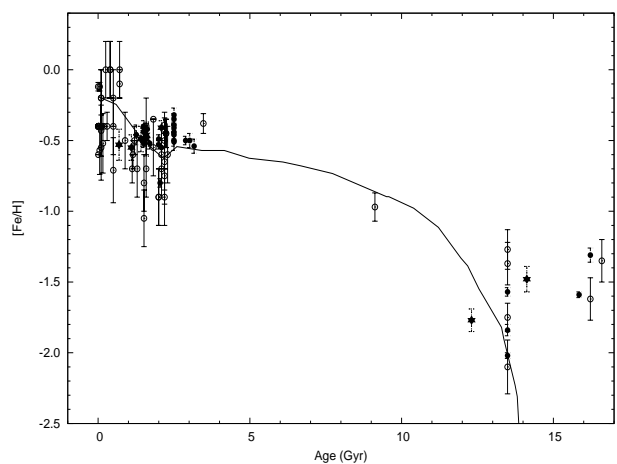}
\caption{ The Age-Metallicity Relation for the LMC. The points
with error bars are 101 LMC star clusters which have age and metallicity measurements in the literature and in the present study. 
Star symbols are the points from the present study, closed circles from G06 and open circles from Harris \& Zaritsky (2009), respectively.
Pagel \& Tautvaisiene (1999) burst model is shown as a solid line.
 }
\label{age}
\end{figure*}

The age-metallicity relation (AMR) is a record of the progressive chemical enrichment of the star-forming 
local interstellar medium during the evolution of the galaxy, and so provides useful clues about the star
formation and chemical evolution history of the local environment (Carraro et al. 1998).
All existing AMR studies suffer from uncertainties due to either inhomogeneous datasets or
photometry based abundances. In Figure \ref{age}, we make an updated AMR using homogeneous set of data
using the CaT method for [Fe/H] measurement (36 clusters, 8 from the
present work (star symbols) \& 28 from G06 (closed circles)).
We have also plotted for comparison, the data from the sample of Harris \& Zaritsky (2009) (open circles)
which are not in common with either the present work or the work done by G06.
We can easily see there is less scatter in [Fe/H] values for the clusters using the CaT method.

The essential features of the AMR have not changed (see also Harris \& Zaritsky 2009): there is still a long age gap, which is also a metallicity gap, especially
if one only looks at the CaT data. 
Prior to the age gap, the LMC formed a number of low-metallicity populous clusters, analogous to the MW’s 
globular cluster population. Following the age gap, cluster formation resumed in the LMC, and proceeded 
roughly continuously to the present day. The clusters formed following the age gap have less dispersion 
in metallicity and are signiﬁcantly more enriched than the ancient cluster population, and there is
some evidence for further enrichment with time for ages $<1$ Gyr.

C05 have obtained stellar metallicities for almost 400 stars in the bar of the LMC, also using CaT lines. 
Their obtained AMR shows a similar behavior to that of the clusters for the oldest population. However, while 
for the clusters the metallicity has increased over the last 2 Gyr, this has not happened in the bar. 
Recently, Carrera et al. (2008) have examined the AMR for field stellar populations, based on CaT
spectroscopy of individual red giants in four LMC fields. They find that the disk AMR is similar to that
of the LMC star clusters: there was rapid initial enrichment to [Fe/H]$\sim-0.8$ by about 10 Gyr ago. Following
this, the metallicity increased only slowly until 3 Gyr ago, when it began a more rapid increase to its present-day 
value of [Fe/H]$\sim-0.2$. However, the lack of objects  with ages between 3 and 10 Gyr is not observed in the field population. 
Their reconstructed SFH (for their inner disk fields) indicates there was a
relatively  quiescent epoch between 5 and
10 Gyr ago but some star formation occurred. These conclusions differ in detail from those derived from the AMR of the central region of the LMC
(C05), but the observed AMRs are entirely consistent.

Carrera et al. (2008) also showed that
the bar AMR differs from that of the disk in the last 3 Gyr: while in the disk the
metallicity has increased during this time, in the bar it has remained approximately constant.
This would be in agreement with the prediction that a
typical bar instability pushes the gas of the disk towards the center (Sellwood \& Wilkinson
1993), so the infall gas is expected to be previously pre-enriched   by the disk chemical
evolution. This would be also the case in the scenario suggested by Bekki \& Chiba (2005),
in which the bar would have formed from disk material as a consequence of tidal interactions
between the LMC, the SMC and the MW about 5 Gyr ago, the epoch  when the
bar AMR differed from that of the disk. Also, a bar is expected to destroy any metallicity
gradient within a certain radius, as is observed (e.g. G06).
Harris \& Zaritsky (2009) recently found that the chemical evolution history
for the LMC is qualitatively similar to the 
one derived for the SMC (see Harris \& Zaritsky 2004). In both galaxies, the
mean metallicity remained at an almost constant, low value, from $\sim$10 Gyr
until about 4 Gyr ago when the metallicity began a steady ascent 
to the present-day value. In both galaxies, the AMR is consistent with
the bursting enrichment model proposed by Pagel \& Tautvaisiene (1999) (P09).

\subsection{Metallicity Gradient}

The galactic distribution of elemental abundances is central to study the chemical evolution of
the galaxy. The relative abundances of the elements are sensitive to the star formation
history, the number of massive stars, the ratio of SN Ia to II,
the importance of AGB stars, the relative yield of the elements, and the exchange
of matter between the galactic disk, bar and halo through infall or ejection. 
Measurements of elemental abundances throughout the galaxy  provide vital input to model the
formation and evolution of the galaxy.
From observations of clusters, O91 and
Santos Jr. et al. (1999) found no evidence for the presence of a radial metallicity gradient in
the LMC. The only evidence for a radial metallicity gradient in the LMC cluster system was
reported by Kontizas, Kontizas, \& Michalitsiano (1993) based on six outer LMC clusters 
($\ge$ 8 kpc). Hill et al. (1995) were the first to report evidences of a gradient in
the field population from high-resolution spectroscopy, using a sample of nine stars in the
bar and in the disk. They found that the bar is on average 0.3 dex more metal-rich than
the disk population, but the disk stars studied are located within a radius of $2^\circ$. 
Subsequent work by Cioni \& Habing (2003) also
implies a decrease in metallicity while moving away from the bar within radius of $6^\circ$. 
Finally, an outward radial gradient of decreasing metallicity
was also found by Alves (2004) from infrared CMDs using the 2MASS survey.
Recently Carrera et al. (2008) concluded that the lower
mean metallicity in the outermost field is a consequence of it having a lower fraction of 
young stars, which are more metal-rich. They found that there is a change in the age composition of the disk
population beyond a certain radius, while the chemical enrichment history seems to be shared by all fields.

\begin{figure*}
\centering
\hbox{
\includegraphics[height=8cm,width=8cm,angle=-90]{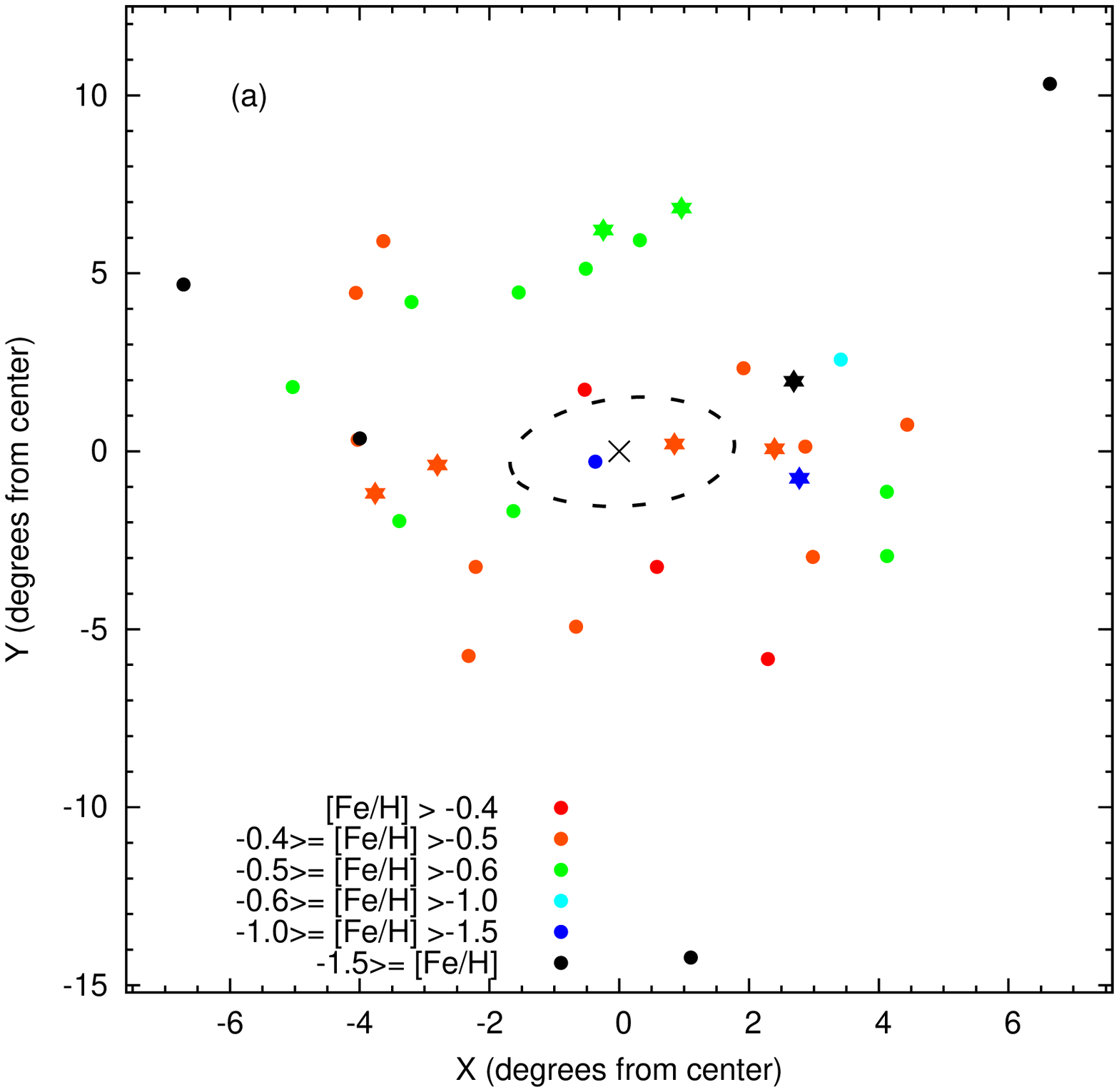}
\includegraphics[height=8cm,width=8cm,angle=-90]{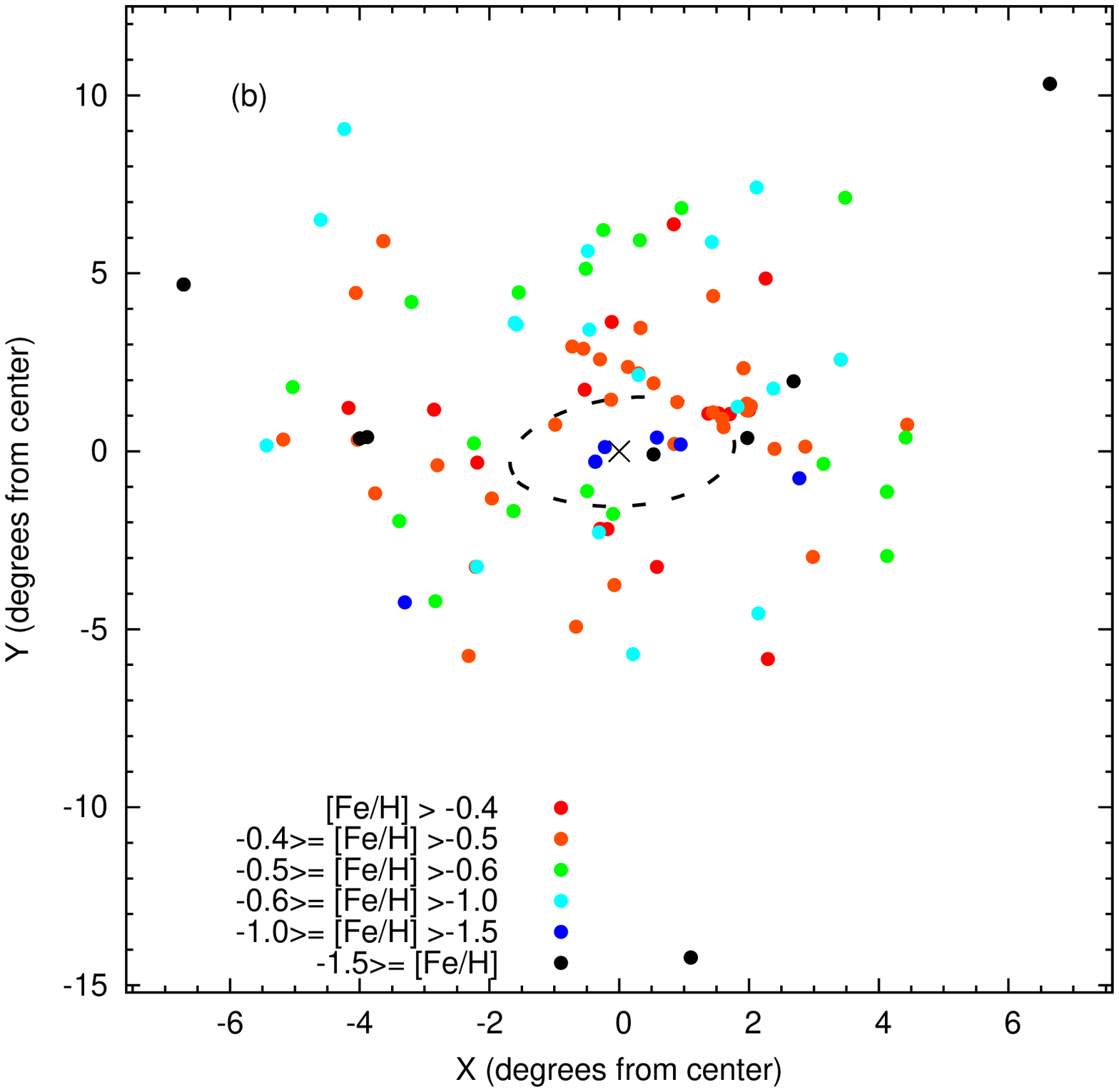}
}
\caption{ 
Positions on the sky and derived metallicities for our target clusters. Metallicity bins are given in the lower left corner of the plot. The adopted LMC
center is marked with the cross, and the dashed line roughly outlines the bar.
Panel (a) represents  clusters with [Fe/H] determined using CaT method, symbols are as same as in Figure \ref{age}. 
Panel (b) represents the sample of all the 101 clusters.
 }
\label{xy}
\end{figure*}

It is necessary to further investigate the presence, or absence, of an abundance gradient in the LMC. 
We have used the current sample of 101 clusters in this regard.
Positions on the sky for each cluster are shown in Figure \ref{xy}
along with the metallicity bin into which each cluster falls represented by the color of the plotting symbol. 
We have used the same color scheme as used by G06. Figure \ref{xy}(a) represents clusters with [Fe/H]
determined using CaT method. Star and dot symbols represent data from the 
present work and from G06 respectively.
Figure \ref{xy}(b) represents the sample of all 101 clusters.
The adopted center of the LMC ($\alpha =  5^h 27^m 36^s , \delta=-69^\circ 52^\prime  12^{\prime\prime}$ ; 
van der Marel et al. 2002) is marked by the cross and the dashed oval represents the $2^\circ$ 
near-infrared isopeth from van der Marel (2001), which roughly outlines the location of the LMC bar. Conversion 
from right ascension and declination  to Cartesian coordinates was performed using a zenithal
equidistant projection (e.g., van der Marel \& Cioni 2001, their eqs. [1]-[4]).

\begin{figure*}
\centering
\includegraphics[height=8cm,width=8cm,angle=-90]{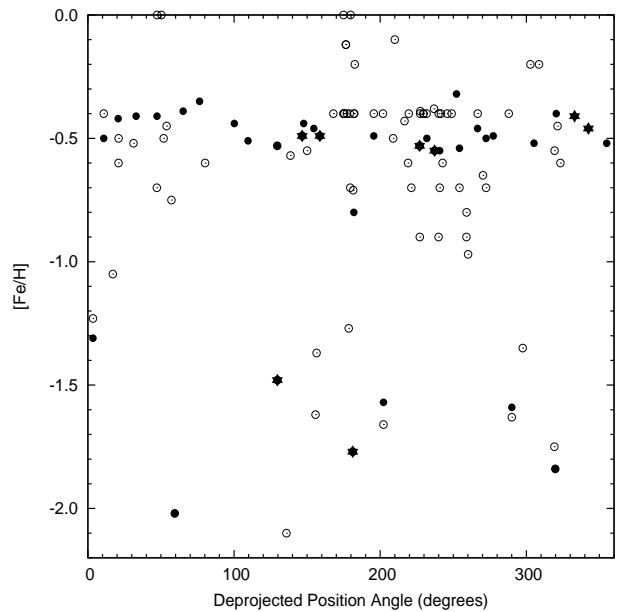}
\caption{ 
Metallicities of our target clusters as a function of de-projected position angle. 
We have used the LMC geometry of van der Marel \& Cioni (2001)
to correct for projection effects. This plot illustrates that there is no apparent
relation between position angle and metallicity in the LMC. 
Symbols are as same as in Figure \ref{age}.
 }
\label{dpotn}
\end{figure*}

\begin{figure*}
\centering
\includegraphics[height=8cm,width=8cm,angle=-90]{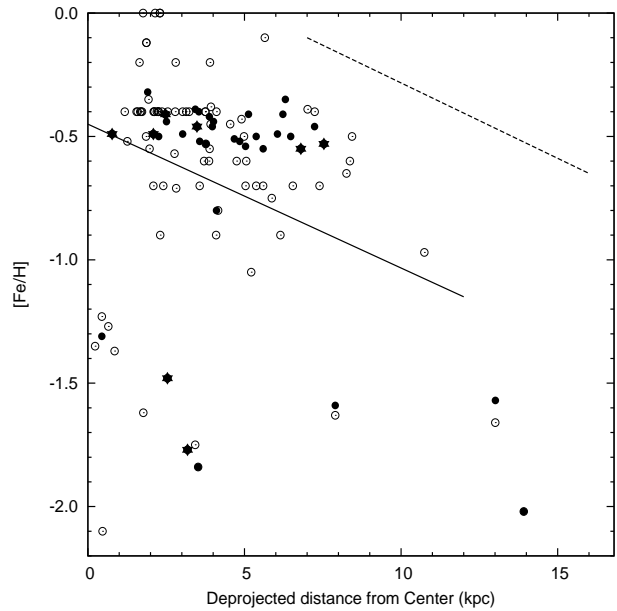}
\caption{ 
Cluster metallicities as a function of de-projected distance (in kpc 
from the center of the LMC). We have assumed a distance of $(m-M_0) = 18.5$.
Over-plotted are the metallicity gradients observed in the MW open clusters
(dashed line; Friel et al. 2002) and M33 (solid line; Tiede et al. 2004), which 
help to further illustrate that the LMC’s cluster system lacks the metallicity 
gradient typically seen in spiral galaxies. The lack of a  gradient is likely caused
by the presence of the central bar (Zaritsky et al. 1994). 
Symbols are as same as in Figure \ref{age}.
 }
\label{ddis}
\end{figure*}

In Figures \ref{dpotn} and \ref{ddis} we further explore the metallicity-position relationship for LMC 
clusters by plotting metallicity as a function of de-projected position angle 
and radial distance (in kpc), respectively. 
The symbols are the same as in Figure \ref{age}. We can very easily see the tight distribution of
the clusters with CaT-based                  [Fe/H] determination.
This is an important factor in constraining the SFH of the galaxy.
We have corrected for projection effects by adopting $34^\circ.7$ as the inclination 
and $122^\circ.5$ for the position angle of the line of nodes of
the LMC (van der Marel \& Cioni 2001 and G06). In this rotated coordinate system, 
a cluster with a position angle of zero lies along
the line of nodes, and angles increase counterclockwise (see G06); Radial distances were converted from angular 
separation to kpc by assuming an LMC distance of $(m - M_0)$ = 18.5 ($\sim$50 kpc); 
at this distance, $1^\circ$ is $\sim870$ pc.  
In Figure \ref{ddis} we have over-plotted both the MW open cluster metallicity gradient from Friel et al. (2002; dashed line) and the
M33 gradient from Tiede et al. (2004; solid line). Neither of these disk abundance gradients resembles what we see among the LMC
clusters, which shows no evidence for a gradient.

Combined, these three figures illustrate that,
similar to what was found by O91 (and Geisler et al. 2003, G06), there
is no [Fe/H] gradient in terms of either position angle or radial
distance for the higher metallicity clusters in our sample. 
It is well known that a number of
metal-poor clusters ([Fe/H] $\leq -1.5$) exist in the inner portions
of the LMC (e.g., O91), suggesting that neither the metal-rich   
nor the metal-poor    clusters exhibit a metallicity gradient.
In the case of the LMC, the presence of a strong bar component may have diluted the metallicity
gradient originally present in the star clusters, leading to a cluster population that is well mixed (see G06). 

Geisler et al. (2009) also found that the cluster distribution is similar to what has been 
found for red giant stars in the bar, which indicates that the bar and the intermediate age clusters have similar
 star formation histories. This is in good agreement with recent theoretical models that suggest the bar and
 intermediate age clusters formed as a result of a close encounter with the SMC. Our findings also confirm
 previous results which show that the LMC lacks the metallicity gradient typically seen in non-barred
 spiral galaxies, suggesting that the bar is driving the mixing of stellar populations in the LMC.

\subsection{Kinematics}

The kinematical properties of the LMC provide important clues to its structure.
They have been obtained from many tracers such as HI gas (e.g., Rohlfs et al. 1984; Luks \& Rohlfs 1992; Kim et al. 1998),
star clusters (Freeman, Illingworth \& Oemler 1983; Schommer et al. 1992), planetary nebulae (Meatheringham et al. 1988),
HII regions and supergiants (Feitzinger, Schmidt-Kaler \& Isserstedt 1977), and carbon stars 
(Kunkel et al. 1997; Graff et al. 2000; Alves \& Nelson 2000; van der Marel et al. 2002).
All these studies imply that the LMC is kinematically cold, and must therefore to first  approximation be a disk system
(van der Marel 2004).
G06            also found their cluster sample has motions consistent with that of a single rotating disk system, 
with no indication of halo kinematics. 

To study the kinematics of the LMC, we followed the approach of G06 and Schommer et al. (1992) using star clusters as a tracers.
To characterize the rotation of their clusters, Schommer et al.
(1992) fit an equation of the form

$ V(\theta) = \pm V_m \{ [{\rm tan}(\theta-\theta_0)~{\rm sec}~i]^2 +1 \}^{-0.5} + V_{sys} $\\
to their radial velocity data using a least-squares technique to
derive the systemic velocity $(V_{sys})$, the amplitude of the rotation
velocity $(V_m)$, and the orientation of the line of nodes; they
adopted an inclination of $27^\circ$. Their best-fit parameters give a
rotation amplitude and dispersion consistent with the LMC clusters having disk like 
kinematics, with no indications of the existence of a pressure-supported halo.

\begin{figure*}
\centering
\includegraphics[height=8cm,width=8cm,angle=-90]{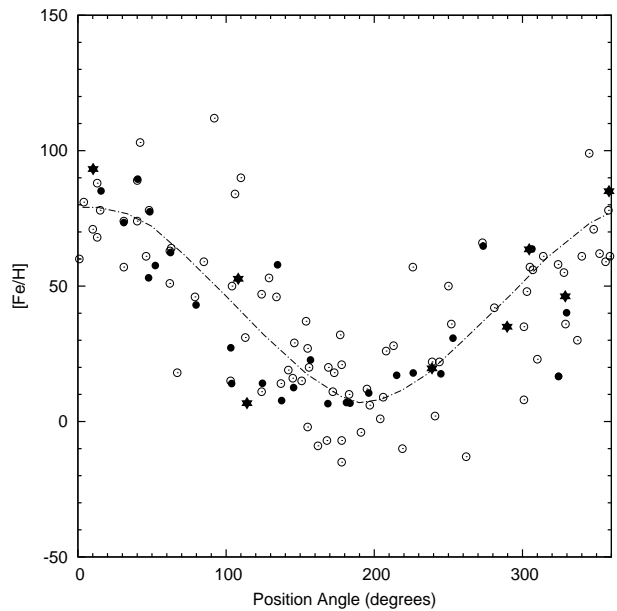}
\caption{ 
Galactocentric radial velocities as a function of position angle on
the sky for the clusters in our sample (stars), from G06 (filled circles) as well as those from
Schommer et al. (1992; open circles). 
Rotation curve solution number 3 from
Schommer et al. (1992) is over-plotted as the dashed line, showing that all  data
sets are consistent with circular rotation. 
 }
\label{potn}
\end{figure*}

In Figure \ref{potn} we have plotted galactocentric radial velocity
versus position angle on the sky for our sample (8 from present work + 28 from G06), along with velocity 
data for all clusters listed in Schommer et al. (1992). To be
consistent with the approach of Schommer et al. (1992), we have
adopted the galactocentric velocity corrections given by Feitzinger
\& Weiss (1979). In addition, for this Figure only, we have adopted
their LMC center ($\alpha= 5^h 20^m 40^s , \delta = -69^\circ 14^\prime 10^{\prime\prime}$ [J2000])
for use in calculating the position angles of our clusters (see also G06). 
We have used the standard astronomical convention in which north has a
position angle of zero and angles increase to the east (G06); 
Data from Schommer et al. (1992) are plotted as open circles, present data
are plotted as star symbols and data from G06 is given by solid points.
Over-plotted on this figure (dashed line) is the rotation curve solution 
number 3 from Schommer et al. (1992). 
Derived velocities for the present eight target clusters show that their motions
are consistent with the findings of Schommer et al. (1992) in that
the LMC cluster system exhibits disk-like kinematics that are very
similar to the HI disk and has no obvious signature of a stellar halo.
Carrera (2008) argued that failure to find evidence of a hot stellar halo is related to   a low
contrast of the halo population with respect to that of the disk, even at the large galactocentric
radius of their outermost field. Curiously, both old, metal-poor
clusters as well as intermediate age, metal-rich clusters
share the same disk rotation.

\section{Summary and Conclusion}

In this paper, we have studied the properties of CMDs, age, metallicity and 
radial velocities of eight LMC massive  clusters. 
Several of these clusters have no modern metallicity determination;
out of eight, four have a spectroscopically determined metallicity value given in O91 with a larger error.
We have used on-average 10 cluster members to determine the mean metallicity which shows a scatter of only    0.05 - 0.12 dex.
To derive the metallicity, we obtained medium-resolution spectra at the near-infrared CaT
using the FORS2 multiobject spectrograph at the 8.2-meter VLT/UT1 with high signal-to-noise ($20 \leq S/N \leq 100$)
and then used the empirically developed, simply calibrated method which can make
an efficient estimate of metallicity ([Fe/H]) for individual RGB
stars using the strength of the CaT lines. 
We also studied in detail the CMDs of all eight clusters based on photometric data
taken from FORS2 + VLT under excellent atmospheric conditions (seeing $<$ 0.7 arcsec) and determined their ages, making use of existing deep HST data for
several of the more crowded fields.
The strong CaT lines are also used to determine the radial velocity of each target.          

We have used the derived age, metallicity and radial velocities of all eight clusters  
to study the age metallicity distribution, their relationship and significance, their distribution in 
the LMC and the kinematics of the LMC. We have confirmed    the efficiency
and efficacy of the CaT method in deriving [Fe/H]
with much less scatter than previous studies, yielding more conclusive results.  
These data can further be used to calibrate the RGB slope-metallicity-age relation 
based on our homogeneous spectroscopic and infrared data. The main results of this paper are as follows:

\begin{itemize}

\item

We have studied the CMDs and reported the age of all eight clusters.
Six have ages between 
0.8-2.2 Gyr and the other two, ie. NGC 1754 and NGC 1786, are very old.

\item

We have derived a spectroscopically determined metallicity for all eight clusters, with
comparatively less scatter than previous studies using the CaT method.
Out of eight, four have no previously spectroscopically determined metallicity value and the
rest have large errors in their previously determined values. The cluster Hodge 2 has no previously reported metallicity value.

\item

The metallicity of the six intermediate age clusters, having a mean age 1.5 Gyr,
 is $-0.49$ with a scatter of 
0.04, in excellent agreement with the previous finding of G06, which shows a very tight distribution
with mean metallicity of $-0.49$ ($\sigma$ = 0.09).
 Similarly, Mucciarelli et al. (2008) used            high resolution spectra of the giants in 4 LMC globular clusters
and found a mean [Fe/H] = $-0.38\pm0.09$. 
This agrees well with the theoretical work of Bekki et al. (2004), which indicates that a
close encounter between the LMC and SMC caused not only the restart of cluster formation in the
LMC but the generation of the central bar (mean [Fe/H] = $-0.37\pm0.15$; C05).

\item

We find that there is no evidence of a metallicity gradient in the LMC,
 in contrast with the stellar
populations of the MW and M33, which show that the metallicity decreases as galactocentric distance increases.
The LMC's stellar Bar may be the responsible factor for the dilution of any kind of gradient in the LMC.
This result is  in good agreement with previous work done by G06.

\item
 
We find that our derived cluster velocities are in good agreement with the results of Schommer et al. (1992) and G06, confirming
that the LMC cluster system exhibits disk-like rotation with no clusters appearing to have halo kinematics.

\end{itemize}

\ack
JB is supported by FONDECYT \#1080086 and MIDEPLAN ICM Nucleus P07-021-F. RK is supported by Fondecyt \#1080154.
D.G. gratefully acknowledges support from the Chilean  {\sl Centro de Astrof\'\i sica} FONDAP No. 15010003
and from the Chilean Centro de Excelencia en Astrof\'\i sica y Tecnolog\'\i as Afines (CATA). S.S. acknowledges support from Comitee Mixto ESO-GOBIERNO DE CHILE and from GEMINI-CONICYT FUND \#32090002. SS received partial support from Center of Excellence in Astrophysics and Associated Technologies BASAL CATA PFB-06.
We would like to thank an anonymous referee for comments that helped to improve the clarity of this manuscript.
The data used in this paper have been obtained
with 2.5m du Pont  at Las Campanas Observatory and with FORS2/VLT at the
ESO La Silla and Paranal Observatories under observing program: 078.D-0428.  
 We thank the staff at  Las Campanas, la Silla and Paranal observatories for their assistance during the observations. 
This research has made use of the SIMBAD
database, operated at CDS, Strasbourg, France.

\begin{table*}[t]\tabcolsep=0.1pt\small
\begin{center}
\caption{The log of the imaging observations.
}
\label{log}
\begin{tabular}{l@{\hspace{5pt}}l@{\hspace{5pt}}l@{\hspace{15pt}}lllrc}
\hline
\multicolumn{1}{l}{Object}{\hspace{5pt}}&
\multicolumn{1}{l}{$\alpha_{2000}$}&
\multicolumn{1}{l}{$\delta_{2000}$}{\hspace{5pt}}&
\multicolumn{1}{r}{Exptime V}{\hspace{3pt}}&
\multicolumn{1}{c}{Exptime R}{\hspace{3pt}}\\
\hline
NGC\,1795& 04:59:46 & $-$69:48:06 & 15s  &20s \\
NGC\,1754& 04:54:17 & $-$70:26:30 & 15s  &20s     \\
NGC\,1786& 04:59:06 & $-$67:44:42 & 15s  &20s \\   
Hodge\,2 & 05:17:49 & $-$69:38:40 & 15s  & 20s \\
NGC\,1900& 05:19:09 & $-$63:01:24 & 15s  &20s \\
SL\,509  & 05:29:48 & $-$63:39:00 & 15s  &20s  \\
SL\,817  & 06:00:38 & $-$70:04:07 & 15s  & 20s \\
SL\,862  & 06:13:27 & $-$70:41:42 & 15s  &20s  \\
\hline
\end{tabular}
\end{center}
\end{table*}

\begin{deluxetable}{ r r r r r r c c c c c c c c }
\tablewidth{0pt}
\tablehead{
\colhead{Star ID} &\colhead{ $\alpha_{2000}$ } &\colhead{ $\delta_{2000}$} & \colhead{RV}  &\colhead{ $ \sigma_{\rm RV}$} &\colhead{ $V-V_{HB}$} &\colhead{ S/N} &\colhead{ $W$}  &\colhead{ $W^\prime$} &\colhead{ $\sigma _{W^\prime}$}  &\colhead{ [Fe/H]}&\colhead{ $\sigma_{\rm[Fe/H]}$}&\colhead{ Comment} \\
        &\colhead{(h:m:s)}&\colhead{(d:m:s)} &\colhead{ (km $s^{-1}$)}&\colhead{ (km $s^{-1}$)}&\colhead{ (mag)} & & &&&\colhead{(dex)}&\colhead{ (dex)}&
}
\tabletypesize{\tiny}
\tablecaption{Position and measured values for all RGB stars.\label{data} }

\startdata
NGC1795-01 &  04:59:44.64 & $-$69:49:22.4&  260.6 & 7.7&  $-$1.92&  52&  8.58 & 7.18& 0.06& $-$0.37& 0.11& m \\
NGC1795-02 &  04:59:47.93 & $-$69:49:16.3&  246.2 & 7.7&  $-$0.65&  40&  7.63 & 7.15& 0.05& $-$0.38& 0.11& m \\
NGC1795-07 &  04:59:48.96 & $-$69:48:48.6&  259.5 & 7.6&  $-$1.09&  44&  7.62 & 6.82& 0.01& $-$0.50& 0.10& m \\
NGC1795-08 &  04:59:58.88 & $-$69:48:41.8&  244.0 & 8.1&  $-$0.63&  44&  6.81 & 6.35& 0.09& $-$0.67& 0.11& m \\
NGC1795-11 &  04:59:49.41 & $-$69:48:24.1&  260.0 & 7.9&  $-$1.49&  45&  8.29 & 7.20& 0.06& $-$0.36& 0.11& m \\
NGC1795-17 &  04:59:44.36 & $-$69:47:45.9&  261.0 & 7.6&  $-$1.07&  44&  7.60 & 6.82& 0.01& $-$0.50& 0.10& m \\
NGC1795-18 &  04:59:53.13 & $-$69:47:39.5&  254.1 & 7.7&  $-$1.76&  47&  8.09 & 6.81& 0.01& $-$0.50& 0.10& m \\
NGC1795-19 &  04:59:50.27 & $-$69:47:36.3&  261.8 & 7.5&  $-$2.00&  46&  8.71 & 7.25& 0.07& $-$0.34& 0.11& m \\
NGC1795-20 &  04:59:33.18 & $-$69:47:27.6&  247.4 & 7.6&  $-$1.57&  58&  7.76 & 6.61& 0.04& $-$0.57& 0.10& m \\
NGC1795-03 &  05:00:00.20 & $-$69:49:14.2&   28.2 & 7.4&  $ $0.06&  58&  7.08 & 7.13& 0.05& $-$0.39& 0.11& n \\
NGC1795-04 &  04:59:51.54 & $-$69:49:05.9&  202.7 & 7.7&  $-$1.77&  60&  7.74 & 6.44& 0.07& $-$0.63& 0.10& n \\
NGC1795-05 &  04:59:58.99 & $-$69:48:58.3&  229.0 & 7.5&  $-$0.88&  54&  7.60 & 6.95& 0.02& $-$0.45& 0.10& n \\
NGC1795-06 &  05:00:04.00 & $-$69:48:54.4&  227.4 & 7.8&  $-$0.99&  48&  7.79 & 7.07& 0.04& $-$0.41& 0.11& n \\
NGC1795-09 &  04:59:49.92 & $-$69:48:33.8&  217.3 & 8.8&  $-$1.34&  51&  7.81 & 6.83& 0.01& $-$0.49& 0.10& n \\
NGC1795-10 &  04:59:46.68 & $-$69:48:31.7&  148.2 & 8.2&  $-$1.52&  62&  6.71 & 5.60& 0.22& $-$0.94& 0.15& n \\
NGC1795-13 &  04:59:44.57 & $-$69:48:11.9&  189.3 & 9.0&  $-$1.41&  49&  8.10 & 7.07& 0.04& $-$0.41& 0.11& n \\
NGC1795-14 &  04:59:47.36 & $-$69:48:05.0&  232.6 & 8.4&  $-$1.38&  56&  7.40 & 6.39& 0.08& $-$0.65& 0.11& n \\
NGC1795-15 &  04:59:51.65 & $-$69:47:57.1&  130.7 & 9.1&  $-$1.90&  52&  8.32 & 6.94& 0.02& $-$0.46& 0.10& n \\
NGC1795-16 &  04:59:45.55 & $-$69:47:51.7&  125.0 & 9.4&  $-$2.24&  54&  7.84 & 6.20& 0.12& $-$0.72& 0.11& n \\
NGC1795-21 &  04:59:54.39 & $-$69:47:22.5&  232.9 & 7.6&  $-$0.98&  54&  7.48 & 6.77& 0.01& $-$0.52& 0.10& n \\
NGC1795-23 &  04:59:34.06 & $-$69:47:08.5&  220.0 & 7.6&  $-$2.01&  50&  8.38 & 6.92& 0.01& $-$0.46& 0.10& n \\
NGC1795-24 &  04:59:28.41 & $-$69:47:04.9&  255.7 & 7.5&  $-$1.22&  43&  8.00 & 7.11& 0.05& $-$0.39& 0.11& n \\
NGC1795-26 &  04:59:43.62 & $-$69:46:51.2&  221.5 & 7.7&  $-$1.98&  47&  8.67 & 7.22& 0.07& $-$0.35& 0.11& n \\
NGC1795-27 &  04:59:34.73 & $-$69:46:45.1&  254.3 & 7.6&  $-$0.50&  36&  6.77 & 6.40& 0.08& $-$0.65& 0.11& n \\
NGC1795-28 &  04:59:35.86 & $-$69:46:36.8&  258.1 & 7.7&  $-$2.03&  43&  8.72 & 7.24& 0.07& $-$0.35& 0.11& n \\
NGC1795-29 &  04:59:48.68 & $-$69:46:32.5&  211.0 & 7.6&  $-$1.31&  53&  7.48 & 6.52& 0.06& $-$0.60& 0.10& n \\
NGC1795-36 &  04:59:38.20 & $-$69:45:49.3&  218.1 & 7.7&  $-$1.05&  48&  7.09 & 6.32& 0.10& $-$0.68& 0.11& n \\
NGC1795-37 &  04:59:46.35 & $-$69:45:44.6&  260.7 & 7.6&  $-$1.37&  52&  8.06 & 7.06& 0.04& $-$0.41& 0.11& n \\
NGC1795-38 &  04:59:42.55 & $-$69:45:38.2&  225.7 & 7.6&  $-$2.20&  48&  9.68 & 8.07& 0.22& $-$0.04& 0.17& n \\
NGC1754-11 &  04:54:19.41 & $-$70:26:46.7&  228.9 & 7.8&  $-$1.36&  46&  4.62 & 3.63& 0.35& $-$1.65& 0.14& m \\
NGC1754-12 &  04:54:19.68 & $-$70:26:40.9&  236.6 & 7.7&  $-$2.65&  46&  5.98 & 4.05& 0.26& $-$1.50& 0.11& m \\
NGC1754-13 &  04:54:17.58 & $-$70:26:36.2&  246.0 & 7.6&  $-$2.61&  58&  5.84 & 3.94& 0.28& $-$1.54& 0.12& m \\
NGC1754-14 &  04:54:21.93 & $-$70:26:28.3&  227.7 & 7.6&  $-$1.87&  64&  5.77 & 4.41& 0.18& $-$1.37& 0.10& m \\
NGC1754-15 &  04:54:19.78 & $-$70:26:21.5&  231.5 & 7.5&  $-$1.70&  59&  5.45 & 4.21& 0.22& $-$1.44& 0.10& m \\
NGC1754-16 &  04:54:14.57 & $-$70:26:15.4&  233.6 & 7.7&  $-$1.20&  39&  5.36 & 4.48& 0.16& $-$1.34& 0.09& m \\
NGC1754-17 &  04:54:21.45 & $-$70:26:12.5&  231.4 & 7.7&  $-$1.55&  43&  5.11 & 3.98& 0.27& $-$1.52& 0.12& m \\
NGC1754-19 &  04:54:27.43 & $-$70:25:58.1&  237.0 & 7.6&  $-$1.65&  62&  5.24 & 4.03& 0.26& $-$1.51& 0.11& m \\
NGC1754-02 &  04:54:02.34 & $-$70:27:41.8&  210.3 & 7.7&  $-$1.08&  46&  7.15 & 6.36& 0.24& $-$0.66& 0.13& n \\
NGC1754-05 &  04:54:37.18 & $-$70:27:25.2&  248.8 & 7.5&  $-$1.42&  49&  6.86 & 5.83& 0.12& $-$0.86& 0.10& n \\
NGC1754-06 &  04:54:26.82 & $-$70:27:18.7&  263.0 & 7.6&  $-$2.68&  41&  8.16 & 6.21& 0.20& $-$0.72& 0.12& n \\
NGC1754-08 &  04:54:20.76 & $-$70:27:06.1&  223.2 & 7.6&  $-$2.24&  53&  8.04 & 6.41& 0.25& $-$0.65& 0.13& n \\
NGC1754-10 &  04:54:12.80 & $-$70:26:54.2&  248.2 & 7.5&  $-$1.35&  46&  6.96 & 5.97& 0.15& $-$0.80& 0.11& n \\
NGC1754-20 &  04:54:31.48 & $-$70:25:52.3&   38.3 & 7.5&  $-$1.92&  59&  5.40 & 4.00& 0.27& $-$1.52& 0.12& n \\
NGC1754-21 &  04:54:20.83 & $-$70:25:45.8&  252.3 & 7.6&  $-$1.45&  57&  6.14 & 5.08& 0.04& $-$1.13& 0.08& n \\
NGC1754-24 &  04:54:34.96 & $-$70:25:27.1&  260.3 & 7.4&  $-$1.84&  58&  5.52 & 4.17& 0.23& $-$1.46& 0.11& n \\
NGC1754-25 &  04:54:14.16 & $-$70:25:19.5&  226.9 & 7.5&  $-$3.17&  47&  8.68 & 6.37& 0.24& $-$0.66& 0.13& n \\
NGC1754-27 &  04:54:15.40 & $-$70:25:07.3&  233.4 & 7.4&  $-$2.17&  55&  7.92 & 6.34& 0.23& $-$0.67& 0.13& n \\
NGC1754-29 &  04:54:09.36 & $-$70:24:56.9&  284.8 & 7.5&  $-$3.00&  52&  7.62 & 5.43& 0.04& $-$1.00& 0.08& n \\
NGC1754-30 &  04:54:13.52 & $-$70:24:48.6&  259.2 & 7.5&  $-$2.46&  49&  8.18 & 6.39& 0.24& $-$0.65& 0.13& n \\
NGC1754-32 &  04:54:29.56 & $-$70:24:36.0&  260.5 & 7.6&  $-$1.78&  60&  7.15 & 5.85& 0.13& $-$0.85& 0.10& n \\
NGC1754-34 &  04:54:03.51 & $-$70:24:24.1&  260.5 & 7.4&  $-$1.88&  54&  7.49 & 6.12& 0.19& $-$0.75& 0.11& n \\
NGC1754-35 &  04:54:11.56 & $-$70:24:18.0&  260.1 & 7.6&  $-$2.16&  46&  7.97 & 6.39& 0.24& $-$0.65& 0.13& n \\
NGC1754-37 &  04:54:13.22 & $-$70:24:05.8&  266.1 & 7.6&  $-$1.76&  58&  7.53 & 6.25& 0.21& $-$0.71& 0.12& n \\
NGC1786-03 &  04:59:04.08 & $-$67:45:47.9&  264.1 & 7.6&  $-$2.06& 112&  4.88 & 3.38& 0.20& $-$1.74& 0.09& m \\
NGC1786-06 &  04:59:09.12 & $-$67:45:29.2&  265.4 & 7.6&  $-$1.28&  70&  4.14 & 3.21& 0.23& $-$1.81& 0.10& m \\
NGC1786-09 &  04:59:09.43 & $-$67:45:09.4&  272.3 & 8.0&  $-$1.90& 103&  4.89 & 3.50& 0.18& $-$1.70& 0.09& m \\
NGC1786-18 &  04:59:07.00 & $-$67:44:17.5&  268.1 & 7.4&  $-$1.55&  53&  4.19 & 3.06& 0.26& $-$1.86& 0.11& m \\
NGC1786-19 &  04:59:06.71 & $-$67:44:12.1&  273.4 & 7.5&  $-$2.30&  85&  5.36 & 3.68& 0.14& $-$1.63& 0.08& m \\
NGC1786-20 &  04:59:08.31 & $-$67:44:03.1&  273.4 & 7.6&  $-$1.77&  80&  4.56 & 3.27& 0.22& $-$1.78& 0.10& m \\
NGC1786-22 &  04:59:08.15 & $-$67:43:48.7&  279.4 & 8.2&  $-$1.29&  47&  4.00 & 3.06& 0.26& $-$1.86& 0.11& m \\
NGC1786-05 &  04:59:07.04 & $-$67:45:33.8&  243.1 & 7.8&  $-$0.62&  41&  5.34 & 4.89& 0.09& $-$1.20& 0.08& n \\
NGC1786-08 &  04:59:03.32 & $-$67:45:16.6&  124.9 &10.3&  $-$2.98&  91&  5.64 & 3.46& 0.18& $-$1.71& 0.09& n \\
NGC1786-10 &  04:59:08.23 & $-$67:45:03.2&  222.3 & 8.5&  $-$2.03&  84&  4.62 & 3.14& 0.25& $-$1.83& 0.10& n \\
NGC1786-11 &  04:59:08.31 & $-$67:44:56.8&  219.3 & 8.4&  $-$2.78&  95&  5.86 & 3.83& 0.11& $-$1.58& 0.07& n \\
NGC1786-12 &  04:59:05.93 & $-$67:44:52.4&  151.6 & 9.1&  $-$2.98&  89&  5.65 & 3.48& 0.18& $-$1.71& 0.09& n \\
NGC1786-13 &  04:59:04.82 & $-$67:44:45.3&  173.0 & 8.6&  $-$2.84&  96&  5.69 & 3.62& 0.15& $-$1.66& 0.08& n \\
NGC1786-14 &  04:59:09.90 & $-$67:44:38.4&  204.4 & 8.5&  $-$2.13&  82&  4.90 & 3.34& 0.21& $-$1.76& 0.09& n \\
NGC1786-15 &  04:59:10.72 & $-$67:44:32.3&  118.9 & 8.9&  $-$2.87&  96&  5.64 & 3.54& 0.17& $-$1.68& 0.09& n \\
NGC1786-16 &  04:59:05.92 & $-$67:44:27.6&  162.1 & 8.9&  $-$2.66&  96&  5.50 & 3.56& 0.17& $-$1.68& 0.08& n \\
NGC1786-17 &  04:59:05.98 & $-$67:44:22.6&  175.8 & 9.3&  $-$1.38&  61&  4.70 & 3.69& 0.14& $-$1.63& 0.08& n \\
NGC1786-21 &  04:59:04.49 & $-$67:43:56.6&  262.4 & 7.3&  $-$0.80&  38&  5.35 & 4.76& 0.07& $-$1.24& 0.08& n \\
NGC1786-24 &  04:58:54.33 & $-$67:43:38.6&  217.9 & 7.6&  $-$1.00&  49&  7.35 & 6.62& 0.42& $-$0.57& 0.18& n \\
NGC1786-25 &  04:58:52.60 & $-$67:43:33.9&  259.5 & 7.4&  $-$1.93&  64&  4.82 & 3.41& 0.19& $-$1.73& 0.09& n \\
NGC1786-27 &  04:59:04.93 & $-$67:43:21.0&  289.1 & 7.7&  $-$0.86&  41&  7.30 & 6.67& 0.43& $-$0.55& 0.19& n \\
NGC1786-32 &  04:59:00.39 & $-$67:42:48.2&  248.6 & 7.5&  $-$0.79&  44&  6.89 & 6.31& 0.36& $-$0.68& 0.16& n \\
Hodge02-05 &  05:17:54.04 & $-$69:39:33.8&  281.6 & 7.4&  $-$1.84&  39&  7.82 & 6.48& 0.07& $-$0.62& 0.10& m \\
Hodge02-07 &  05:17:49.49 & $-$69:39:22.0&  260.1 & 7.5&  $-$1.30&  48&  7.71 & 6.76& 0.01& $-$0.52& 0.10& m \\
Hodge02-10 &  05:17:43.56 & $-$69:39:01.4&  273.3 & 7.7&  $-$1.16&  42&  7.12 & 6.28& 0.10& $-$0.69& 0.10& m \\
Hodge02-11 &  05:17:50.19 & $-$69:38:57.1&  253.3 & 7.5&  $-$1.99&  56&  8.68 & 7.23& 0.08& $-$0.35& 0.11& m \\
Hodge02-20 &  05:17:48.66 & $-$69:38:03.1&  275.1 & 7.6&  $-$2.80&  53&  9.06 & 7.01& 0.03& $-$0.43& 0.10& m \\
Hodge02-22 &  05:17:39.87 & $-$69:37:53.4&  263.5 & 7.5&  $-$1.97&  55&  8.67 & 7.23& 0.08& $-$0.35& 0.11& m \\
Hodge02-23 &  05:17:40.98 & $-$69:37:45.5&  257.0 & 7.5&  $-$1.42&  54&  8.00 & 6.96& 0.02& $-$0.45& 0.10& m \\
Hodge02-01 &  05:17:41.43 & $-$69:39:54.7&  292.0 & 7.6&  $-$2.27&  48&  9.13 & 7.47& 0.12& $-$0.26& 0.12& n \\
Hodge02-02 &  05:17:36.04 & $-$69:39:51.5&  197.7 & 7.5&  $-$2.08&  56&  9.09 & 7.57& 0.14& $-$0.22& 0.12& n \\
Hodge02-03 &  05:17:38.40 & $-$69:39:45.0&  145.3 & 8.6&  $-$1.17&  54&  8.02 & 7.17& 0.06& $-$0.37& 0.11& n \\
Hodge02-04 &  05:17:51.68 & $-$69:39:40.7&  288.7 & 7.6&  $-$1.00&  49&  8.18 & 7.45& 0.12& $-$0.27& 0.12& n \\
Hodge02-06 &  05:17:41.80 & $-$69:39:27.4&  267.3 & 7.6&  $-$0.64&  38&  8.00 & 7.53& 0.13& $-$0.24& 0.12& n \\
Hodge02-08 &  05:17:41.62 & $-$69:39:13.7&  126.2 &10.7&  $-$0.72&  42&  7.05 & 6.53& 0.06& $-$0.60& 0.10& n \\
Hodge02-09 &  05:17:48.16 & $-$69:39:08.3&  178.8 & 8.9&  $-$1.77&  47&  8.81 & 7.52& 0.13& $-$0.24& 0.12& n \\
Hodge02-13 &  05:17:44.53 & $-$69:38:43.4&  293.9 & 7.6&  $-$2.03&  50&  8.26 & 6.78& 0.01& $-$0.51& 0.10& n \\
Hodge02-14 &  05:17:50.76 & $-$69:38:37.7&  211.8 & 8.5&  $-$0.94&  56&  7.14 & 6.46& 0.07& $-$0.63& 0.10& n \\
Hodge02-15 &  05:17:51.50 & $-$69:38:31.2&  231.8 & 8.2&  $-$1.49&  43&  8.75 & 7.66& 0.16& $-$0.19& 0.13& n \\
Hodge02-16 &  05:17:49.96 & $-$69:38:27.6&  247.9 & 7.9&  $-$2.05&  65&  6.39 & 4.90& 0.36& $-$1.19& 0.15& n \\
Hodge02-17 &  05:17:49.06 & $-$69:38:24.0&  255.4 & 7.6&  $-$1.94&  46&  9.05 & 7.64& 0.15& $-$0.20& 0.12& n \\
Hodge02-18 &  05:17:51.89 & $-$69:38:16.4&  219.3 & 7.5&  $-$1.66&  61&  6.81 & 5.60& 0.23& $-$0.94& 0.12& n \\
Hodge02-21 &  05:17:56.14 & $-$69:37:59.2&  238.3 & 7.5&  $-$2.05&  73&  6.76 & 5.26& 0.30& $-$1.06& 0.13& n \\
Hodge02-24 &  05:17:36.30 & $-$69:37:40.4&  244.5 & 7.5&  $-$1.14&  40&  8.45 & 7.62& 0.15& $-$0.21& 0.12& n \\
Hodge02-25 &  05:17:43.88 & $-$69:37:34.3&  279.1 & 7.5&  $-$1.49&  57&  6.08 & 4.99& 0.35& $-$1.16& 0.15& n \\
Hodge02-27 &  05:17:35.55 & $-$69:37:22.1&  287.0 & 7.5&  $-$1.13&  47&  8.16 & 7.34& 0.10& $-$0.31& 0.11& n \\
Hodge02-29 &  05:17:40.16 & $-$69:37:08.4&  298.9 & 7.6&  $-$1.94&  48&  9.21 & 7.79& 0.18& $-$0.15& 0.13& n \\
Hodge02-31 &  05:17:37.64 & $-$69:36:57.9&  254.9 & 7.5&  $-$0.95&  42&  7.11 & 6.41& 0.08& $-$0.64& 0.10& n \\
Hodge02-36 &  05:17:27.05 & $-$69:36:28.1&  264.2 & 7.5&  $-$0.97&  35&  7.51 & 6.80& 0.01& $-$0.50& 0.10& n \\
Hodge02-38 &  05:17:29.64 & $-$69:36:16.2&  256.2 & 7.4&  $-$1.06&  43&  7.74 & 6.97& 0.03& $-$0.44& 0.10& n \\
NGC1900-06 &  05:19:00.29 & $-$63:02:11.8&  311.8 & 8.5&  $-$0.72&  51&  7.22 & 6.70& 0.08& $-$0.54& 0.10& m \\
NGC1900-07 &  05:19:14.61 & $-$63:02:01.7&  312.6 & 8.1&  $-$0.21&  42&  6.49 & 6.34& 0.01& $-$0.67& 0.09& m \\
NGC1900-10 &  05:19:09.13 & $-$63:01:45.5&  291.8 &10.8&  $-$1.41&  57&  7.56 & 6.53& 0.04& $-$0.60& 0.10& m \\
NGC1900-11 &  05:19:07.56 & $-$63:01:38.3&  309.2 & 8.4&  $-$1.01&  48&  7.67 & 6.93& 0.14& $-$0.46& 0.11& m \\
NGC1900-17 &  05:19:08.82 & $-$63:01:03.0&  321.3 & 8.3&  $-$0.61&  51&  7.29 & 6.85& 0.12& $-$0.49& 0.11& m \\
NGC1900-19 &  05:19:08.97 & $-$63:00:50.0&  302.8 & 8.2&  $-$0.67&  57&  6.72 & 6.23& 0.03& $-$0.71& 0.09& m \\
NGC1900-21 &  05:19:07.38 & $-$63:00:35.6&  300.2 & 8.3&  $-$0.42&  43&  7.51 & 7.21& 0.20& $-$0.36& 0.13& m \\
NGC1900-24 &  05:19:04.35 & $-$63:00:17.6&  296.0 & 8.3&  $-$0.56&  54&  6.87 & 6.46& 0.03& $-$0.63& 0.10& m \\
NGC1900-26 &  05:19:12.46 & $-$63:00:02.9&  300.6 & 8.3&  $-$1.38&  64&  7.74 & 6.74& 0.09& $-$0.53& 0.11& m \\
NGC1900-27 &  05:19:09.75 & $-$62:59:58.2&  303.3 & 8.2&  $-$0.79&  50&  7.44 & 6.86& 0.12& $-$0.48& 0.11& m \\
NGC1900-04 &  05:19:17.94 & $-$63:02:21.1&  333.3 & 8.1&  $-$0.06&  46&  7.16 & 7.11& 0.18& $-$0.39& 0.12& n \\
NGC1900-08 &  05:19:11.22 & $-$63:01:57.4&  -39.4 & 8.3&  $-$0.69&  49&  7.08 & 6.58& 0.06& $-$0.58& 0.10& n \\
NGC1900-09 &  05:19:11.47 & $-$63:01:49.4&  222.9 &13.0&  $-$0.61&  52&  6.90 & 6.45& 0.03& $-$0.63& 0.10& n \\
NGC1900-12 &  05:19:09.78 & $-$63:01:30.4&  316.0 & 8.8&  $-$0.91&  39&  6.59 & 5.93& 0.09& $-$0.82& 0.10& n \\
NGC1900-15 &  05:19:06.08 & $-$63:01:16.0&  228.2 &13.8&  $-$1.56&  48&  7.92 & 6.78& 0.10& $-$0.51& 0.11& n \\
NGC1900-16 &  05:19:08.29 & $-$63:01:06.2&  -32.6 & 9.0&  $-$1.07&  61&  5.80 & 5.02& 0.30& $-$1.15& 0.13& n \\
NGC1900-23 &  05:18:58.42 & $-$63:00:24.1&  325.3 & 8.1&  $-$1.46&  74&  6.01 & 4.94& 0.32& $-$1.18& 0.14& n \\
NGC1900-25 &  05:19:05.56 & $-$63:00:11.9&   14.3 & 8.0&  $-$1.93&  69&  6.30 & 4.89& 0.33& $-$1.20& 0.14& n \\
SL509-02   &  05:29:51.15 & $-$63:40:13.8&  321.8 & 7.5&  $-$0.03&  40&  7.11 & 7.09& 0.09& $-$0.40& 0.11& m \\
SL509-05   &  05:29:43.32 & $-$63:39:50.8&  309.0 & 7.6&  $ $0.06&  44&  6.85 & 6.89& 0.05& $-$0.47& 0.10& m \\
SL509-06   &  05:29:46.03 & $-$63:39:47.2&  307.0 & 7.6&  $ $0.09&  35&  6.68 & 6.74& 0.03& $-$0.53& 0.10& m \\
SL509-08   &  05:29:46.39 & $-$63:39:33.5&  328.2 & 7.6&  $-$0.24&  47&  6.92 & 6.75& 0.03& $-$0.52& 0.10& m \\
SL509-09   &  05:29:52.91 & $-$63:39:28.1&  326.5 & 7.4&  $-$0.32&  39&  6.74 & 6.50& 0.02& $-$0.61& 0.10& m \\
SL509-10   &  05:29:53.58 & $-$63:39:21.6&  311.3 & 7.6&  $-$0.58&  54&  7.14 & 6.72& 0.02& $-$0.54& 0.10& m \\
SL509-13   &  05:29:52.05 & $-$63:39:03.2&  324.8 & 7.6&  $-$1.38&  55&  7.82 & 6.82& 0.04& $-$0.50& 0.10& m \\
SL509-14   &  05:29:49.59 & $-$63:38:55.7&  317.8 & 7.7&  $-$1.33&  56&  7.57 & 6.60& 0.01& $-$0.58& 0.10& m \\
SL509-15   &  05:29:49.76 & $-$63:38:50.3&  314.9 & 7.7&  $-$0.66&  55&  6.92 & 6.44& 0.03& $-$0.63& 0.10& m \\
SL509-16   &  05:29:53.11 & $-$63:38:44.5&  310.2 & 7.6&  $-$1.23&  55&  8.05 & 7.15& 0.11& $-$0.38& 0.11& m \\
SL509-19   &  05:29:47.79 & $-$63:38:31.6&  318.1 & 7.6&  $-$0.90&  52&  7.26 & 6.60& 0.01& $-$0.58& 0.10& m \\
SL509-23   &  05:29:55.08 & $-$63:38:07.8&  329.0 & 7.5&  $-$1.34&  49&  7.22 & 6.24& 0.07& $-$0.71& 0.10& m \\
SL509-26   &  05:29:49.46 & $-$63:37:46.6&  302.2 & 7.6&  $-$0.74&  57&  7.01 & 6.47& 0.03& $-$0.62& 0.10& m \\
SL509-01   &  05:29:37.28 & $-$63:40:18.5&  301.9 & 7.7&  $ $0.08&  38&  5.86 & 5.92& 0.14& $-$0.82& 0.10& n \\
SL509-11   &  05:29:51.97 & $-$63:39:14.4&  319.4 & 7.6&  $-$1.27&  55&  8.38 & 7.45& 0.16& $-$0.27& 0.12& n \\
SL509-18   &  05:29:47.83 & $-$63:38:33.0&  297.1 & 7.6&  $-$1.01&  51&  7.46 & 6.72& 0.02& $-$0.53& 0.10& n \\
SL509-21   &  05:29:46.26 & $-$63:38:20.0&  313.6 & 7.6&  $-$0.29&  52&  6.14 & 5.93& 0.13& $-$0.82& 0.10& n \\
SL509-22   &  05:30:03.41 & $-$63:38:11.0&   52.8 & 7.1&  $-$0.36&  51&  5.46 & 5.20& 0.28& $-$1.08& 0.13& n \\
SL509-25   &  05:30:00.51 & $-$63:37:53.8&  312.5 & 7.6&  $-$0.93&  49&  7.65 & 6.97& 0.07& $-$0.44& 0.11& n \\
SL509-27   &  05:29:34.11 & $-$63:37:39.0&  334.4 & 7.7&  $-$0.30&  46&  6.42 & 6.20& 0.08& $-$0.72& 0.10& n \\
SL509-28   &  05:29:42.26 & $-$63:37:31.8&  304.6 & 7.7&  $ $0.14&  35&  6.47 & 6.57& 0.01& $-$0.59& 0.10& n \\
SL509-29   &  05:29:55.16 & $-$63:37:24.6&  295.3 & 7.5&  $-$1.15&  48&  8.20 & 7.36& 0.15& $-$0.30& 0.12& n \\
SL509-30   &  05:29:47.72 & $-$63:37:12.7&  293.6 & 7.5&  $-$0.13&  45&  6.76 & 6.66& 0.01& $-$0.55& 0.10& n \\
SL509-34   &  05:29:30.97 & $-$63:36:45.4&  346.2 & 7.7&  $ $0.03&  37&  6.78 & 6.80& 0.04& $-$0.50& 0.10& n \\
SL509-35   &  05:29:56.90 & $-$63:36:38.2&  328.9 & 7.6&  $ $0.01&  41&  6.56 & 6.57& 0.01& $-$0.59& 0.10& n \\
SL817-09   &  06:00:49.13 & $-$70:04:42.6&  283.2 & 7.5&  $ $0.26&  52&  7.01 & 7.20& 0.08& $-$0.36& 0.11& m \\
SL817-10   &  06:00:37.22 & $-$70:04:38.3&  285.3 & 7.4&  $-$0.31&  66&  7.22 & 6.99& 0.04& $-$0.43& 0.10& m \\
SL817-11   &  06:00:35.82 & $-$70:04:27.5&  282.6 & 7.6&  $-$1.12&  58&  7.92 & 7.10& 0.06& $-$0.39& 0.11& m \\
SL817-12   &  06:00:40.30 & $-$70:04:23.9&  284.5 & 7.6&  $ $0.01&  58&  7.23 & 7.24& 0.09& $-$0.35& 0.11& m \\
SL817-13   &  06:00:38.11 & $-$70:04:17.1&  287.3 & 7.6&  $-$1.60&  56&  8.22 & 7.05& 0.05& $-$0.41& 0.11& m \\
SL817-14   &  06:00:39.02 & $-$70:04:13.1&  272.8 & 7.6&  $-$1.09&  62&  7.54 & 6.74& 0.01& $-$0.53& 0.10& m \\
SL817-15   &  06:00:38.93 & $-$70:04:08.8&  276.3 & 7.6&  $-$0.34&  55&  7.45 & 7.20& 0.08& $-$0.36& 0.11& m \\
SL817-16   &  06:00:36.91 & $-$70:04:04.8&  284.3 & 7.5&  $-$1.20&  57&  7.79 & 6.91& 0.02& $-$0.46& 0.10& m \\
SL817-18   &  06:00:38.30 & $-$70:03:51.1&  277.4 & 7.6&  $-$0.55&  57&  7.47 & 7.07& 0.05& $-$0.41& 0.11& m \\
SL817-21   &  06:00:35.74 & $-$70:03:31.7&  277.7 & 7.5&  $-$0.57&  56&  7.54 & 7.12& 0.06& $-$0.39& 0.11& m \\
SL817-01   &  06:00:25.79 & $-$70:05:33.4&  277.2 & 7.6&  $-$0.50&  67&  5.93 & 5.56& 0.24& $-$0.95& 0.12& n \\
SL817-02   &  06:00:45.40 & $-$70:05:25.8&   51.5 & 7.5&  $-$0.62&  63&  5.82 & 5.37& 0.28& $-$1.02& 0.13& n \\
SL817-03   &  06:00:34.47 & $-$70:05:21.5&  325.5 & 7.6&  $-$0.23&  58&  7.02 & 6.86& 0.01& $-$0.48& 0.10& n \\
SL817-04   &  06:00:44.07 & $-$70:05:12.5&  246.0 & 7.7&  $ $0.44&  40&  6.22 & 6.54& 0.05& $-$0.60& 0.10& n \\
SL817-05   &  06:00:29.38 & $-$70:05:06.7&  254.1 & 7.6&  $ $0.64&  41&  6.28 & 6.75& 0.01& $-$0.52& 0.10& n \\
SL817-06   &  06:00:22.00 & $-$70:04:58.8&  288.5 & 7.8&  $ $0.09&  50&  6.54 & 6.61& 0.04& $-$0.57& 0.10& n \\
SL817-08   &  06:00:43.52 & $-$70:04:49.8&   42.3 & 7.5&  $ $0.81&  52&  5.92 & 6.51& 0.06& $-$0.61& 0.10& n \\
SL817-17   &  06:00:39.04 & $-$70:03:58.0&  266.2 & 7.6&  $ $0.38&  49&  5.99 & 6.27& 0.10& $-$0.70& 0.10& n \\
SL817-19   &  06:00:34.37 & $-$70:03:43.2&  274.1 & 7.6&  $ $0.26&  46&  7.33 & 7.52& 0.14& $-$0.24& 0.12& n \\
SL817-22   &  06:00:40.63 & $-$70:03:25.9&  250.8 & 7.7&  $ $0.26&  47&  6.63 & 6.82& 0.01& $-$0.50& 0.10& n \\
SL817-26   &  06:00:48.89 & $-$70:03:02.2&  224.6 & 7.4&  $ $0.16&  55&  6.87 & 6.98& 0.04& $-$0.44& 0.10& n \\
SL817-27   &  06:00:22.66 & $-$70:02:56.0&  273.3 & 7.5&  $ $0.24&  44&  7.00 & 7.17& 0.07& $-$0.37& 0.11& n \\
SL817-30   &  06:00:48.56 & $-$70:02:36.3&  212.6 & 7.4&  $ $0.59&  47&  6.74 & 7.17& 0.07& $-$0.37& 0.11& n \\
SL817-34   &  06:00:28.67 & $-$70:02:07.4&  285.7 & 7.6&  $ $0.70&  40&  5.94 & 6.45& 0.07& $-$0.63& 0.10& n \\
SL817-35   &  06:00:29.58 & $-$70:01:59.9&  286.7 & 7.5&  $ $0.15&  45&  6.92 & 7.03& 0.05& $-$0.42& 0.10& n \\
SL817-36   &  06:00:29.60 & $-$70:01:58.1&  288.3 & 7.5&  $-$0.07&  53&  6.58 & 6.53& 0.05& $-$0.60& 0.10& n \\
SL862-04   &  06:13:33.79 & $-$70:42:40.7&  226.9 & 7.1&  $-$0.17&  58&  6.93 & 6.81& 0.01& $-$0.50& 0.10& m \\
SL862-06   &  06:13:29.20 & $-$70:42:31.7&  233.1 & 7.1&  $-$0.07&  59&  6.91 & 6.86& 0.02& $-$0.48& 0.10& m \\
SL862-09   &  06:13:26.25 & $-$70:42:11.5&  242.5 & 7.3&  $-$0.34&  59&  7.35 & 7.10& 0.07& $-$0.39& 0.11& m \\
SL862-10   &  06:13:34.48 & $-$70:42:05.4&  237.0 & 7.1&  $ $0.05&  50&  7.01 & 7.05& 0.06& $-$0.41& 0.11& m \\
SL862-11   &  06:13:33.62 & $-$70:42:00.4&  235.6 & 7.2&  $-$0.25&  51&  6.87 & 6.69& 0.01& $-$0.55& 0.10& m \\
SL862-15   &  06:13:25.57 & $-$70:41:41.3&  242.2 & 7.0&  $ $0.18&  40&  6.46 & 6.59& 0.03& $-$0.58& 0.10& m \\
SL862-17   &  06:13:27.10 & $-$70:41:29.0&  243.5 & 7.2&  $-$0.12&  50&  7.14 & 7.05& 0.06& $-$0.41& 0.11& m \\
SL862-18   &  06:13:26.47 & $-$70:41:25.4&  245.2 & 7.1&  $-$0.11&  42&  7.17 & 7.09& 0.07& $-$0.40& 0.11& m \\
SL862-19   &  06:13:33.76 & $-$70:41:19.3&  235.0 & 7.2&  $-$0.11&  50&  7.28 & 7.20& 0.09& $-$0.36& 0.11& m \\
SL862-20   &  06:13:38.76 & $-$70:41:13.6&  226.3 & 7.2&  $-$0.46&  49&  6.96 & 6.63& 0.02& $-$0.57& 0.10& m \\
SL862-21   &  06:13:29.94 & $-$70:41:07.4&  245.5 & 7.1&  $ $0.14&  38&  6.93 & 7.03& 0.06& $-$0.42& 0.11& m \\
SL862-22   &  06:13:19.81 & $-$70:41:01.0&  243.1 & 7.2&  $-$0.67&  59&  7.12 & 6.63& 0.02& $-$0.56& 0.10& m \\
SL862-24   &  06:13:29.45 & $-$70:40:47.6&  238.5 & 7.2&  $-$1.40&  52&  8.00 & 6.98& 0.05& $-$0.44& 0.10& m \\
SL862-01   &  06:13:33.92 & $-$70:43:00.1&  268.3 & 7.2&  $-$0.93&  61&  7.38 & 6.70& 0.01& $-$0.54& 0.10& n \\
SL862-02   &  06:13:16.75 & $-$70:42:57.2&  253.0 & 7.3&  $-$0.54&  52&  7.39 & 7.00& 0.05& $-$0.43& 0.10& n \\
SL862-03   &  06:13:04.46 & $-$70:42:48.2&   14.7 & 7.2&  $-$0.25&  57&  5.48 & 5.30& 0.28& $-$1.05& 0.13& n \\
SL862-05   &  06:13:29.66 & $-$70:42:38.5&  240.0 & 7.2&  $-$0.68&  56&  8.10 & 7.61& 0.17& $-$0.21& 0.13& n \\
SL862-07   &  06:13:30.39 & $-$70:42:24.1&  256.5 & 7.2&  $ $0.11&  53&  5.54 & 5.62& 0.22& $-$0.93& 0.12& n \\
SL862-08   &  06:13:17.11 & $-$70:42:17.3&  220.4 & 7.1&  $ $0.20&  47&  7.15 & 7.30& 0.11& $-$0.32& 0.11& n \\
SL862-16   &  06:13:21.83 & $-$70:41:35.9&  241.7 & 7.2&  $-$1.07&  62&  6.98 & 6.20& 0.11& $-$0.72& 0.10& n \\
SL862-23   &  06:13:33.63 & $-$70:40:55.2&  232.1 & 7.3&  $ $0.48&  39&  7.33 & 7.68& 0.18& $-$0.19& 0.13& n \\
SL862-25   &  06:13:12.56 & $-$70:40:42.2&  212.4 & 6.8&  $ $0.49&  43&  4.97 & 5.33& 0.28& $-$1.04& 0.13& n \\
SL862-30   &  06:13:21.38 & $-$70:40:08.0&  285.2 & 7.2&  $ $0.36&  40&  6.31 & 6.57& 0.03& $-$0.59& 0.10& n \\
SL862-31   &  06:13:40.80 & $-$70:39:57.2&  232.3 & 7.0&  $-$0.09&  51&  6.49 & 6.42& 0.06& $-$0.64& 0.10& n \\
SL862-35   &  06:13:16.99 & $-$70:39:33.5&  248.9 & 7.3&  $ $0.01&  46&  6.99 & 7.00& 0.05& $-$0.43& 0.10& n \\
SL862-37   &  06:13:01.91 & $-$70:39:18.4&  256.7 & 7.2&  $ $0.18&  38&  6.66 & 6.79& 0.01& $-$0.51& 0.10& n \\
\enddata
\\

(m) member
(n) non-member
\end{deluxetable}

\begin{table*}
\centering
\tiny
\caption{ Derived LMC cluster properties.}
\begin{tabular}{@{}rrrlccccrrr@{}}
\hline
Cluster &Age & N & RV $\pm ~ \sigma_{\rm RV}$ & RV (O91)& [Fe/H] $\pm ~\sigma_{\rm[Fe/H]}$  & [Fe/H](O91) & [Fe/H] (O$91^b$)& \multicolumn{2}{c}{[Fe/H] $\pm ~\sigma_{\rm[Fe/H]}(photometric)$}\\
        &(Gyr)&   &  (km $s^{-1}$)   &(km $s^{-1}$)  & (dex)                        &  (dex) & (dex) &\multicolumn{2}{c}{(dex)}{Reference} \\
\hline

NGC\,1795&  1.3 & 9 &$255.0\pm6.8 $ &266& $-0.47\pm0.10$ & $-0.23$ &$ -0.33$  &$-$& $-$ \\
NGC\,1754& 14.0 & 8 &$234.1\pm5.4 $ &236& $-1.48\pm0.09$ & $-1.54$ & $-1.28$  &$-1.42\pm0.15$& Olsen et al. (1998) \\
NGC\,1786& 12.3 & 7 &$279.9\pm4.9 $ &264& $-1.77\pm0.08$ & $-1.87$ $(-1.75\pm0.01^a)$ & $-1.59$  &$-2.10\pm0.3$& Brocato et al. (1996) \\
Hodge\,2 &  1.6 & 7 &$266.3\pm9.7 $ & - & $-0.49\pm0.12$ &  $- $ &  $- $  &$-$&$-$ \\
NGC\,1900&  0.8 & 10&$305.0\pm8.3 $ &300& $-0.55\pm0.10$ & $-0.81$ & $-0.70$  &$-$&$-$ \\
SL\,509  &  1.2 & 13&$317.0\pm8.4 $ & - & $-0.54\pm0.09$ & $  -$ & $  -$  &$-0.65\pm0.2$& Piatti et al. (2003) \\
SL\,817  &  2.2 & 10&$281.1\pm4.5 $ & - & $-0.41\pm0.05$ & $  -$ & $  -$  &$-0.35\pm0.2$&  Piatti et al. (2003) \\
SL\,862  &  1.7 & 13&$238.0\pm6.2 $ & - & $-0.47\pm0.07$ & $  -$ & $  -$  &$-0.75\pm0.2$&  Piatti et al. (2003) \\

\hline
\end{tabular}
a: Based on high resolution spectroscopy: Mucciarelli et al. (2009)\\
b: Transformed using equation in text
\label{param}
\end{table*}

\begin{table*}
\centering
\tiny
\caption{ Metallicities of young, intermediate and old clusters.}
\begin{tabular}{@{}ccccccccccc@{}}
\hline
Age range &\multicolumn{3}{c}{CaT data} &  \multicolumn{3}{c}{Literature data} &\multicolumn{3}{c}{all data} \\
(Gyr)     &  [Fe/H] (dex) &     Age(Gyr) & N    &	[Fe/H] (dex) &     Age(Gyr)&  	N   &	[Fe/H] (dex)   &   Age(Gyr) & N     \\
\hline
$<1$ &$-0.55\pm-    $&$ 0.8\pm -  $&   1  &$  -0.34\pm0.18$&$   0.2\pm0.2$&     35 &$   -0.35\pm0.18$&$ 0.2\pm0.3$&    36     \\
$1-3$&$-0.47\pm0.09 $&$2.0\pm0.5  $&  26  &$  -0.64\pm0.18$&$   1.8\pm0.4$&     22 &$   -0.55\pm0.16$&$ 1.9\pm0.5$&    48     \\
$>12$&$-1.65\pm0.22 $&$14.1\pm1.3  $&  7   &$  -1.58\pm0.29$&$  14.5\pm1.4$&     6  &$   -1.62\pm0.26$&$14.3\pm1.3$&    13     \\
\hline
\end{tabular}
\label{stat}
\end{table*}


\clearpage

\end{document}